\documentclass[prd,aps,a4paper,superscriptaddress,twocolumn,nofootinbib]{revtex4}
\usepackage{graphicx}
\usepackage{color}
\usepackage{dcolumn}
\usepackage{bm}
\usepackage{slashed}
\usepackage{amsmath}
\usepackage{latexsym}
\usepackage{amssymb}
\usepackage{mathrsfs}
\usepackage{amsfonts}
\usepackage{url}
\allowdisplaybreaks
\begin{document}
\title{Post-Keplerian waveform model for binary compact object as sources of space-based gravitational wave detector and its implications}

\author{Li-Fang Li}
\affiliation{Institute of Mechanics, Chinese Academy of Sciences, Beijing 100190, China}
\author{Zhoujian Cao
\footnote{corresponding author}} \email[Zhoujian Cao: ]{zjcao@amt.ac.cn}
\affiliation{Institute of Applied Mathematics, Academy of Mathematics and Systems Science, Chinese Academy of Sciences, Beijing 100190, China}
\affiliation{School of Fundamental Physics and Mathematical Sciences, Hangzhou Institute for Advanced Study, UCAS, Hangzhou 310024, China}

\begin{abstract}
Binary compact objects will be among the important sources for the future space-based gravitational wave detectors. Such binary compact objects include stellar massive binary black hole, binary neutron star, binary white dwarf and mixture of these compact objects. Regarding to the relatively low frequency, the gravitational interaction between the two objects of the binary is weak. Post-Newtonian approximation of general relativity is valid. Previous works about the waveform model for such binaries in the literature consider the dynamics for specific situations which involve detailed complicated matter dynamics between the two objects. We here take a different idea. We adopt the trick used in pulsar timing detection. For any gravity theories and any detailed complicated matter dynamics, the motion of the binary can always be described as a post-Keplerian expansion. And a post-Keplerian gravitational waveform model will be reduced. Instead of object masses, spins, matter's equation of state parameters and dynamical parameters beyond general relativity, the involved parameters in our post-Keplerian waveform model are the Keplerian orbit elements and their adiabatic variations. Respect to current planning space-based gravitational wave detectors including LISA, Taiji and Tianqin, we find that the involved waveform model parameters can be well determined. And consequently the detail matter dynamics of the binary can be studied then. For binary with purely gravitational interactions, gravity theory can be constrained well.
\end{abstract}

\maketitle

\section{Introduction}
Binary stars are unique astrophysical laboratories for the study of fundamental physics and cosmology. Such laboratories have been realized through pulsar timing detection of binary pulsar \cite{PhysRevX.11.041050,2004Lorimer} and gravitational wave detection of binary coalescence \cite{2021arXiv211103606T}. But pulsar timing detection and current gravitational wave detection concern two different life stages of the binary system. Pulsar timing detection is about the early inspiral stage. While current gravitational wave detection by ground based detectors is about the stage near merger. In the coming future the space-based gravitational wave detectors will change this situation. Then gravitational wave detection can also provides information about early inspiral stage \cite{PhysRevLett.100.041102,PhysRevD.102.063021}. For example, LISA may resolve $\sim10^4$ double white dwarfs (DWDs) \cite{2001A&A...365..491N,PhysRevLett.128.041101}.

Waveform model is important for gravitational wave data analysis and for extracting the information of the sources. For the merger stage of a binary, extremal general relativity (GR) condition makes numerical relativity calculation necessary to construct the waveform model \cite{PhysRevD.96.044028_SEOBNRE,PhysRevD.101.044049_validSEOBNRE,Liu_2022}. For the early inspiral stage, the gravitational interaction is much weaker. And Newtonian approximation is valid. The binary compact objects including stellar massive binary black hole, binary neutron star, binary white dwarf and mixture of these compact objects belong to such early inspiral binaries for the future space-based gravitational wave detectors. The simplest waveform model for such sources is based on quadrupole formula for monochromatic approximation \cite{PhysRevD.102.063021}. When the GR leading order of inspiral behavior is taken into consideration, the chirp waveform model has been widely used \cite{2002ApJ...575.1030T}. Besides the GR effect, the mass transfer between the two compact objects \cite{Kremer_2017} and the tidal deformation of the two objects \cite{Benacquista_2011,10.1093.mnrasl.slaa183} can also contribute to the chirp signal. Recently, the Ref.~\cite{PhysRevLett.100.041102} used Newtonian approximation to construct a waveform model and focused on the effect of white dwarf interior on the periastron precession. These different waveform models focus on different physical mechanism and introduce different waveform parameters consequently. Such diversity of waveform models make data analysis and source property study hard. In order to catch weak GW signals, ones have to analyze the same data many many times with different specific waveform models. Even worse, some signals may be lost due to the uncomplete analysis. As promising laboratory to test gravity theories, it is interesting and important to construct waveform model for different gravity theories other than GR. Unfortunately such waveform models have not been studied yet.

In contrast to the aforementioned specific waveform models for different physical situations, an unified waveform model will be useful. Such an unified waveform model will definitely facilitate the data analysis for the space-based detectors. With such an unified waveform model, ones can not only catch the signal much easier but also distinguish different physical situations and even constraint different gravity theories. We will propose such an unified waveform model in the current paper.

Since the component objects of a binary in early inspiral stage are well separated, the orbital dynamics can be well described by a perturbation to a Kepler orbit. Such perturbation may come from gravity theory, environment, tidal interaction and mass transfer between the two components. All kinds of effects on the binary dynamics can be included. Consequently, the parameterized post-Keplerian formalism \cite{PhysRevD.45.1840,PhysRevD.93.124061} is a good framework to describe the system. This unified waveform model uses post-Keplerian parameters instead of component masses and other physical parameters. This unified waveform model is consequently valid for any kinds of gravity theories and any detailed complicated matter dynamics. If ones can measure the post-Keplerian parameters accurately, gravity theory and the theory of stellar structure and evolution can be tested and investigated \cite{PhysRevX.11.041050,PhysRevLett.100.041102}. We will show later in the current paper that this is exactly true for the current planning space-based gravitational wave detectors including LISA, Taiji and Tianqin.

The arrangement of the rest of this paper is as following. We will present the post-Keplerian waveform model in the next section. Then we relate the involved post-Keplerian parameters to general relativity (GR) effect, gravity theory effect beyond GR, tidal interaction and the mass transfer effect between the two components of the binary. When we construct such relation we also give guide lines to use post-Keplerian parameters to distinguish these different effects. After that we use Fisher matrix technique to estimate the measurement accuracy of the post-Keplerian parameters based on LISA, Taiji and Tianqin respectively. We find that the involved post-Keplerian parameters can be well determined. And consequently the detail matter dynamics of the binary can be studied then. For binary with purely gravitational interactions, gravity theory can be constrained well. That is to say, equipped with our post-Keplerian waveform model, the future space-based gravitational wave detectors can do good science for compact object binary observations. Throughout this paper we will use units $c=G=1$.
\section{Post-Keplerian waveform model}
The general bound Keplerian orbit is an eccentric orbit. The binary moves along an eccentric orbit and the corresponding gravitational wave form respect to the detector can be written as \cite{2002ApJ...575.1030T,PhysRevD.69.082005,PhysRevLett.100.041102}
\begin{align}
&h=\frac{\sqrt{3}}{2}\sum_{n=1}^\infty[F_+h_n^++F_\times h_n^\times],\label{eq17}\\
&h_n^+(t)=A\{(1+\cos^2\iota)a_n(e)\cos[n\phi(t)]\cos(2\gamma)\nonumber\\
&-(1+\cos^2\iota)b_n(e)\sin[n\phi(t)]\sin(2\gamma)\nonumber\\
&+(\sin^2\iota) c_n(e)\cos[n\phi(t)]\},\\
&h_n^\times(t)=-2A\cos\iota\{b_n(e)\sin[n\phi(t)]\cos(2\gamma)\nonumber\\
&+a_n(e)\cos[n\phi(t)]\sin(2\gamma)\},\\
&a_n(e)=n[J_{n-2}(ne)-2eJ_{n-1}(ne)+(2/n)J_n(ne)\nonumber\\
&+2eJ_{n+1}(ne)-J_{n+2}(ne)],\\
&b_n(e)=n\sqrt{1-e^2}[J_{n-2}(ne)-2J_n(ne)+J_{n+2}(ne)],\\
&c_n(e)=J_n(ne),
\end{align}
where $J_i(x)$ is the $i$-th order Bessel function of the first kind. The amplitude $A$ depends on the gravity theory and the specific dynamics of the binary. $n$ corresponds to higher harmonics excited by the eccentric orbit. In \cite{Liu_2022}, we call it tones to distinguish with the spherical harmonics. $\phi(t)$ is the Doppler shifted orbital phase
\begin{align}
\phi(t)=\phi_0+2\pi ft
\end{align}
with initial phase $\phi_0$.

Perturbations to the above Kepler orbit may result in changing of orbital frequency $f$, inclination angle $\iota$, orbit eccentricity $e$, periastron direction $\gamma$. In all we have adiabatic changing
\begin{align}
&f=f_0+\dot{f}_0t,\\
&\iota=\iota_0+\dot{\iota}_0t,\\
&e=e_0+\dot{e}_0t,\\
&\gamma=\gamma_0+\dot{\gamma}_0t.\label{eq18}
\end{align}
Suitable coordinate choice can eliminate $\gamma_0=0$. The investigation in \cite{PhysRevLett.100.041102} neglected parameters $\dot{\iota}_0$ and $\dot{e}_0$. The authors in \cite{10.1093.mnrasl.slaa183} only considered the changing of $f$ but to second order.

On the first glance, our waveform model has nothing to do with the masses of the two components of the binary. That is because the information of mass and spin are coded in post-Keplerian parameters like $\dot{f}_0$. And just because of this fact, our post-Keplerian waveform model is quite generic for stellar binary sources of space detector.

Besides the above intrinsic parameters, the pattern function $F_{+,\times}$ will introduce three more extrinsic parameters \cite{PhysRevD.57.7089}
\begin{align}
F^+(\alpha,\delta,\psi)&\equiv\frac{1}{2}(1+\sin^2\delta)\cos2\alpha\cos2\psi\nonumber\\
&\,\,\,\,-\sin\delta\sin2\alpha\sin2\psi,\\
F^\times(\alpha,\delta,\psi)&\equiv\frac{1}{2}(1+\sin^2\delta)\cos2\alpha\sin2\psi\nonumber\\
&\,\,\,\,+\sin\delta\sin2\alpha\cos2\psi.
\end{align}
The above pattern functions depends on the source right ascension $\alpha$ and declination $\delta$, and the wave polarization angle $\psi$.

Including both the intrinsic and extrinsic parameters we have 12 parameters $(A,f_0,\dot{f}_0,\iota_0,\dot{\iota}_0,e_0,\dot{e}_0,\dot{\gamma}_0,\phi_0,\alpha,\delta,\psi)$ in all. Source localization can be analyzed as \cite{PhysRevD.76.022003,PhysRevD.101.084053,PhysRevD.102.024089,Ruan2021,PhysRevD.103.064057,PhysRevD.103.103013}. Since we do not care about the source localization in the current work and the source location can not be determined by a single detector, we instead fix $\alpha=\psi=0$ and $\delta=\pi/2$ in the current work. So we need only consider 9 parameters $(A,f_0,\dot{f}_0,\iota_0,\dot{\iota}_0,e_0,\dot{e}_0,\dot{\gamma}_0,\phi_0)$ and the pattern functions become
\begin{align}
F^+=1,F^\times=0.
\end{align}

Eqs.~(\ref{eq17})-(\ref{eq18}) correspond to the post-Keplerian waveform model. The approximated pattern function shown above is only for simplified discussion involved in the next sections. Such simplification does not affect the general conclusion. And the post-Keplerian waveform model can be straight forwardly applied to any realistic situations.
\section{Theoretical predictions on the post-Keplerian parameters}
Assuming the masses of the two components of the binary are $M_{1,2}$, the chirp mass, symmetric mass ratio and total mass are given by
\begin{align}
&\mathcal{M}=\eta^{3/5}M^{2/5},\\
&\eta=\frac{M_1M_2}{M},\\
&M=M_1+M_2.
\end{align}
In order to make the mass transfer between the two compact objects stable till merger, the mass ratio should be limited in $(0,0.24)$ \cite{1988ApJ...332..193V,1999A&A...349L..17H,2010A&A...521A..85Y}.

\subsection{Leading order gravitational radiation reaction effect in GR}
The general relativity contribution to the post-Keplerian parameters can be calculated through post-Newtonian approximation as \cite{PhysRevLett.87.251101,Willems_2007,maggiore2008gravitational,PhysRevD.80.084001,PhysRevD.93.064031,PhysRevD.96.044011}
\begin{align}
&A=\frac{(2\pi \mathcal{M}f_0)^{2/3}}{1-e^2_0}\frac{\mathcal{M}}{D_L},\label{eq5}\\
&\dot{f}_0=\frac{\pi^{8/3}\mathcal{M}^{5/3}f_0^{11/3}}{5(1-e_0^2)^{7/2}}[96+292e_0^2+37e_0^4],\label{eq9}\\
&\dot{\iota}_0=0,\\
&\dot{e}_0=-\frac{2}{3}\frac{\sigma}{\sigma'}\frac{\dot{f}_0}{f_0},\label{eq2}\\
&\sigma=\frac{e_0^{12/19}}{1-e_0^2}\left[1+\frac{121}{304}e_0^2\right]^{870/2299},\\
&\sigma'=\frac{1}{2^{1181/2299}19^{870/2299}}\times\nonumber\\
&\frac{96+292e_0^2+37e_0^4}{e_0^{7/19}(1-e_0^2)^2(304+121e_0^2)^{1429/2299}},
\end{align}
where $D_L$ is the luminosity distance between the source and the detector. $f_0$, $\iota_0$, $e_0$ and $\dot{\gamma}_0$ correspond to the orbital status of the binary. Approximately \cite{PhysRevLett.87.251101}
\begin{align}
f_0&\approx\frac{1}{2\pi}\sqrt{\frac{M}{a^3}},\label{eq10}\\
\dot{\gamma}_0&=6\pi f_0\frac{M}{a(1-e_0^2)}\\
&\approx24\pi^3\frac{a^2f_0^3}{1-e_0^2}\\
&\approx3(2\pi)^{7/3}\frac{M^{2/3}f_0^{5/3}}{1-e_0^2}\label{eq15}
\end{align}
where $a$ is the separation of the two components of the binary, approximately the semi-major axis of the binary orbit. In principle $\iota_0\in(0,\pi)$, $e_0\in(0,1)$. Since $\dot{\iota}_0=0$ and $(\dot{e}_0,\dot{\gamma}_0)$ can be determined by other parameters, there are only six independent parameters involved for pure GR situation.

From the above relations we can find that for a binary with given components mass and source localization
\begin{align}
A&\propto f_0^{2/3},\\
\dot{f}_0&\propto f_0^{11/3},\\
\dot{e}_0&\propto e_0f_0^{8/3},\\
\dot{\gamma}_0&\propto f_0^{5/3}.
\end{align}
This is to say when the system frequency $f_0$ increases, all the above parameters increase correspondingly. If just the eccentricity increases, only $\dot{e}_0$ is affected. But if we only consider $e_0$ takes order $0.1$ which means $e_0\in(0.1,1)$, the difference of $\dot{e}_0$ due to $e_0$ is small compared to the difference due to $f_0$.

If we use the above gravitational radiation reaction to estimate the post-Keplerian parameters, we have typically \cite{Stroeer_2005,2006CQGra..23S.809S,luo2016tianqin}
\begin{align}
&M\approx1M_\odot\approx10^{-6}{\rm s},\label{eq19}\\
&A\approx10^{-20},\label{eq20}\\
&f_0\approx10^{-3}{\rm Hz},\label{eq21}\\
&a\approx1{\rm s},\label{eq22}\\
&\dot{f}_0\approx10^{-18}{\rm Hz}^{2},\label{eq3}\\
&\dot{e}_0\approx10^{-16}{\rm Hz},\\
&\dot{\gamma}_0\approx10^{-6}{\rm Hz}.\label{eq4}
\end{align}
Here Eq.~(\ref{eq19}) is because we care about compact stars including neutron star and white dwarf. Eq.~(\ref{eq20}) corresponds to the verification binaries for space-based detectors. Eq.~(\ref{eq21}) is determined by the sensitive frequency band of space-based detectors. Eqs.~(\ref{eq22})-(\ref{eq4}) are determined by the relations shown in (\ref{eq5})-(\ref{eq15}). The variation time scale is much smaller than the orbit period indicate. This fact means that our post-Keplerian waveform model introduced in the above section is valid quite well for gravitational radiation reaction.

We have two clues to show whether the detected binary is dominated by GR effect or not. The first one is the detected $\dot{\iota}_0$ strongly deviates from 0, and/or $\dot{f}_0$, $\dot{e}_0$ and $\dot{\gamma}_0$ strongly deviate from the predicted values (\ref{eq3})-(\ref{eq4}). The GR effect on the orbit plan precession comes from the compact object spin precession \cite{PhysRevD.49.6274}
\begin{align}
\dot{\vec{L}}=&\frac{1}{a^3}\left[\frac{4M^2_1+3M_1M_2}{2}\vec{\chi}_1+\frac{4M^2_2+3M_1M_2}{2}\vec{\chi}_2\right]\times\vec{L}\nonumber\\
&-\frac{3}{2}\frac{M_1^2M_2^2}{a^3}\left[(\vec{\chi}_2\cdot\hat{L})\vec{\chi}_1+(\vec{\chi}_1\cdot\hat{L})\vec{\chi}_2\right]\times\hat{L}\nonumber\\
&-\frac{32}{5}\frac{\mu^2}{a}\left(\frac{M}{a}\right)^{5/2}\hat{L},
\end{align}
where $\vec{L}$ is the total angular momentum of the binary system, $\vec{\chi}_{1,2}$ are the dimensionless spin vectors of two individual components. For black holes, the dimensionless spin parameter $\chi$ is less than 1 due to the cosmic censorship. For other objects like neutron star and white dwarf, it will be even much smaller, typically $\chi<0.1$. So we can estimate the GR induced
\begin{align}
\dot{\iota}_{\rm GR0}<\chi\left[2\frac{M^2}{a^3}+\frac{3}{2}\frac{M_1^2M_2^2}{a^3M\sqrt{aM}}\right]\sim\chi\frac{M^2}{a^3}\sim10^{-13}
\end{align}
which can be safely neglected compared to the measurement accuracy (check Table.~\ref{tab1}).

\begin{figure}
\begin{tabular}{c}
\includegraphics[width=0.45\textwidth]{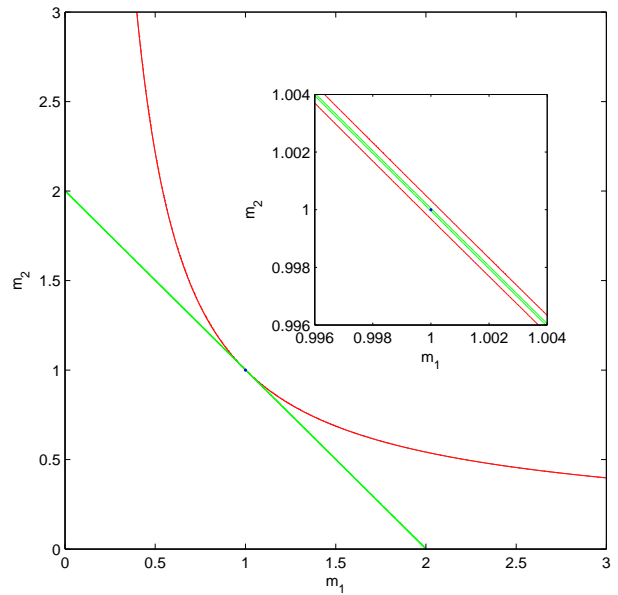}
\end{tabular}
\caption{Plot of companion mass ($M_2$) versus compact object mass ($M_1$) for an example case $M_1=M_2=1M_\odot$. Constraints
derived from the measured post-Keplerian parameters assuming GR are shown pairs of lines with
separation indicating the uncertainty range. The measurement accuracy is taken from Table.~\ref{tab1}. The red lines correspond to $\dot{f}_0$ and the green lines correspond to $\dot{\gamma}_0$. The blue point corresponds to the expected value from general relativity.}\label{fig0}
\end{figure}

The second clue is the relation between $\dot{f}_0$ and $\dot{\gamma}_0$. Using the analysis technique similar to the one in pulsar timing \cite{2015IJMPD..2430018M}, we can draw lines relating $M_1$ and $M_2$ based on the measurement of post-Keplerian parameters $\dot{f}_0$ and $\dot{\gamma}_0$. As an example, we plot such a diagram in Fig.~\ref{fig0} which is similar to Fig.~6 of \cite{2015IJMPD..2430018M} in pulsar timing.
\subsection{Gravitational effect beyond GR}
Using the Brans-Dicke theory as an example \cite{PhysRevD.100.124032,PhysRevD.87.081506}, the gravity theory beyond GR will introduce extra driving terms beside the ones predicted by GR in (\ref{eq9}) and (\ref{eq2})
\begin{align}
&\dot{f}_{\rm 0BD}=\frac{96\pi^2}{5}b\mathcal{M}\eta^{2/5}f_0^3\frac{1+\frac{1}{2}e_0^2}{(1-e_0^2)^{5/2}},\\
&\dot{e}_{\rm 0BD}=-\frac{48\pi^2}{5}b\mathcal{M}\eta^{2/5}f_0^2\frac{e_0}{(1-e_0^2)^{3/2}}.
\end{align}
Here $b$ is a parameter determined by the gravity theory \cite{1989ApJ...346..366W}. When $b=0$ GR is recovered. Periastron precession becomes
\begin{align}
\dot{\gamma}_0&=6\pi f_0\frac{M}{a(1-e_0^2)}\frac{\mathcal{P}}{\mathcal{G}}\label{eq16}
\end{align}
where parameters $\mathcal{P}$ and $\mathcal{G}$ are determined by the gravity theory. For black holes we have
\begin{align}
&\mathcal{G}=1-\frac{\xi}{2},\\
&\mathcal{P}=\mathcal{G}^2,\\
&\xi=\frac{1}{2+\omega_{\rm BD}}.
\end{align}
For neutron star and white dwarf, $b$, $\mathcal{P}$ and $\mathcal{G}$ also depend on the sensitivity of the stars.

If we use parametrized post-newtonian (PPN) formalism to describe the alternative gravity theory other than GR, the factor $\frac{\mathcal{P}}{\mathcal{G}}$ becomes \cite{1987CeMec..40...77S,2013RAA....13.1231Z}
\begin{align}
\frac{\mathcal{P}}{\mathcal{G}}=\frac{2\gamma+2-\beta}{3}.
\end{align}
When the PPN parameters $\gamma=\beta=1$, GR is recovered. In addition, PPN will correct the relation (\ref{eq10}) to
\begin{align}
&f_0\approx\frac{1}{2\pi}\sqrt{\frac{M}{d^3}}\times\nonumber\\
&\left[1-(3\beta+6\gamma-9\eta+6)\frac{M}{d\sqrt{1-e_0^2}}
+(2\gamma-7\eta+4)\frac{M}{d}\right].
\end{align}

If the gravitational constant $G$ is varying, additional driving term beside the one predicted by GR in (\ref{eq9}) comes in \cite{PhysRevLett.61.1151}
\begin{align}
&\dot{f}_{\rm 0VG}=-2\frac{\dot{G}_0}{G_0}f_0,
\end{align}
where $G_0$ is the current value of the gravitational constant.

If dipole radiation exists which results from the equivalence principle violation, the post-Keplerian parameter $\dot{f}_0$ will change to \cite{PhysRevLett.116.241104,PhysRevD.104.084008}
\begin{align}
&\dot{f}_0=\left[1+B\frac{d}{M}\right]\dot{f}_{\rm 0GR},\\
&B=\frac{5}{96}|\Delta\alpha|^2,
\end{align}
where $\Delta\alpha$ describes the difference between the effective scalar couplings
of two objects in the binary. Here we have used $\dot{f}_{\rm 0GR}$ to denote the expression (\ref{eq9}).

If monopole radiation exists, the post-Keplerian parameter $\dot{f}_0$ will change to \cite{Damour_1992,PhysRevX.11.041050,PhysRevD.105.064034}
\begin{align}
&\dot{f}_0=\left[1+Ce_0^2(1+\frac{e_0^2}{4})\right]\dot{f}_{\rm 0GR},
\end{align}
where $C$ depends on specific gravity theory.

Here we just calculated typical gravity theories beyond GR. General gravity theory can be similar related to the post-Keplerian parameters. If only we find a GR effect dominated binary with space-based gravitational wave detector, such binary will be a very good laboratory for gravity theory study.
\subsection{Tidal interactions}
Tidal interactions may affect the properties of the binary compact objects and their evolutions \cite{1985SvAL...11..224S,2010A&A...521A..85Y,Kremer_2015,Kremer_2017}. Since the gravitational wave amplitude just depends on the collective motion of the two components, Eq.~(\ref{eq5}) still holds even tidal interaction appears. But the dynamical variables $\dot{f}_0$, $\dot{\iota}_0$, $\dot{e}_0$ and $\dot{\gamma}_0$ will evolve following different rule. Theoretical models of tidal interaction include equilibrium type \cite{1998ApJ...503..344I,2010ApJ...713..239W,2011ApJ...740L..53P} and dynamical type \cite{2012MNRAS.421..426F,2013MNRAS.430..274F,2014MNRAS.444.3488F}. All of these models predict
\begin{align}
&\dot{f}_0=C_1(f_1-f_0),\label{eq8}\\
&\dot{\iota}_0=-\frac{2\pi f_0}{(1-e_0^2)^2}\times\nonumber\\
&\sum_{i=1}^2k_i\left(\frac{R_i}{a}\right)^5\left(\frac{f_i}{f_0}\right)^2
\left(1+\frac{M_{3-i}}{M_i}\right)\cos\alpha_i\cos\gamma_i,\label{eq14}\\
&\dot{e}_0=-C_2e_0,\label{eq7}\\
&\dot{\gamma}_0=\frac{2\pi f_0}{(1-e_0^2)^2}\times\nonumber\\
&\sum_{i=1}^2\frac{k_i}{M_i}\left(\frac{R_i}{a}\right)^5\left[15M_{3-i}
\frac{1+\frac{3}{2}e_0^2+\frac{1}{8}e_0^4}{(1-e_0^2)^3}+M\left(\frac{f_i}{f_0}\right)^2\right]\label{eq6}
\end{align}
where $R_i$ is the radius of the $i$-th component star, $f_i$ is its self-rotation frequency, and $k_i\in(0,0.75)$ is its quadrupolar apsidal-motion constant. $\alpha_i$ denotes the angle between the $i$-th component's rotation axis and the normal to the orbit plane. $\gamma_i$ denote the angle between the rotation axis and the binary's line of ascending node. The factor $C_1$ in (\ref{eq8}) depends on the orbital separation, the internal structure of the two components. The factor $C_2$ in (\ref{eq7}) depends on the orbital separation, the internal structure of the two components and their rotation.

When $f_0$ is relatively small the contribution of tidal interaction to $\dot{\gamma}_0$ will be larger than the contribution of GR. So if the detected $\dot{\gamma}_0$ is explicitly bigger than the prediction of GR, we can use relation (\ref{eq6}) to analyze the matter property of the binary \cite{PhysRevLett.100.041102}. In contrast, if the detected $\dot{\gamma}_0$ is consistent to the one predicted by GR we are sure the binary is dominated by GR effect.
\subsection{Mass transfer effect}
When the two compact objects approach to each other, the material of one component may be accreted by another one. And the mass transfer happens consequently \cite{doi:10.1146.annurev.aa.15.090177.001015,2021ApJ...923..125W}. The mass transfer effect will affect the post-Keplerian parameters strongly \cite{10.1111.j.1365-2966.2004.07564.x,Gokhale_2007}.

The mass transfer between a close binary can be dived into three types \cite{doi:10.1146.annurev.aa.15.090177.001015}. Case A: If the orbital separation is small enough (less than a few days), the star can fill its Roche lobe during its slow expansion through the main-sequence phase while still burning hydrogen in its core. Case B: If the orbital period is less than about 100 days, but longer than a few days, the star will fill its Roche lobe during the rapid expansion to a red giant with a helium core. If the helium core ignites during this phase and the transfer is interrupted, the mass transfer is case B. Case C: If the orbital period is above 100 days, the star can evolve to the red supergiant phase before it fills its Roche lobe. In this case, the star may have a CO or ONe core. Since we care about the related gravitational wave in mini-Hz band, we only concern Case A mass transfer in the current work.

When no ejected matter leaves a binary system, the mass transfer is said to be conservative. Consequently we have $\dot{J}=\dot{M}=0$ where $J$ is the total orbital angular momentum
\begin{align}
J=M_1M_2\sqrt{\frac{a(1-e^2)}{M}}.
\end{align}
Differentiating the above equation we get
\begin{align}
\frac{\dot{a}}{a}&=2\frac{\dot{J}}{J}-2\frac{\dot{M}_1}{M_1}-2\frac{\dot{M}_2}{M_2}
+\frac{\dot{M}}{M},\\
&=2\dot{M}_1\frac{M_2-M_1}{M_1M_2}.
\end{align}
Together with the Kepler's third law $\frac{1}{a^3}=\frac{4\pi^2f_0^2}{M}$, we get the additional contribution of post-Keplerian parameters due to mass transfer
\begin{align}
&\dot{f}_{\rm 0MTC}=3\frac{\dot{M}_1(M_1-M_2)}{M_1M_2}f_0.
\end{align}

In case of non-conservative mass transfer both mass and angular momentum can be removed from the system. We can use the moment of inertia of the binary
\begin{align}
I=\frac{M_1M_2}{M}a^2
\end{align}
to express the orbital angular momentum
\begin{align}
J=2\pi If_0=2\pi\frac{q}{(1+q)^2}Ma^2f_0.\label{eq13}
\end{align}
For the non-conservative mass transfer part we can approximately assume the mass ratio $q\equiv\frac{M_1}{M_2}$ is a constant. Consequently we have
\begin{align}
\frac{\dot{J}}{J}=\frac{\dot{M}}{M}+2\frac{\dot{a}}{a}+\frac{\dot{f}_0}{f_0}.
\end{align}
Based on the Kepler's third law we have additional relation
\begin{align}
\frac{\dot{M}}{M}=3\frac{\dot{a}}{a}+2\frac{\dot{f}_0}{f_0}.
\end{align}
Combine the above two equations we get
\begin{align}
\frac{\dot{J}}{J}=\frac{5}{3}\frac{\dot{M}}{M}-\frac{1}{3}\frac{\dot{f}_0}{f_0}.\label{eq12}
\end{align}
If we additionally assume isotropic mass loss from the surface of the components we have
\begin{align}
\dot{J}=2\pi\frac{q}{(1+q)^2}\dot{M}a^2f_0.\label{eq11}
\end{align}
Combine Eqs.~(\ref{eq13})-(\ref{eq11}) we have additional contribution of post-Keplerian parameters due to mass transfer
\begin{align}
&\dot{f}_{\rm 0MTN}=-2\frac{\dot{M}}{M}f_0.
\end{align}

Put the conservative and non-conservative mass transfer together we have
\begin{align}
&\dot{f}_{\rm 0MT}=\left[3\frac{(\dot{M}_1-\dot{M})(M_1-M_2)}{M_1M_2}-2\frac{\dot{M}}{M}\right]f_0.
\end{align}

Mass transfer and tidal interaction also affect the stellar oscillations \cite{10.3389.fspas.2021.663026}. If both gravitational wave and stellar oscillations are observed, the combined information can be used to infer the property of the binary.

Typically $\dot{f}_{\rm 0MT}$ is comparable to $\dot{f}_{\rm 0GR}$. So if the mass transfer does happen we expect the detected $\dot{f}_{\rm 0}$ is different to the one predicted by GR. For such case we can subtract $\dot{f}_{\rm 0GR}$ from the detected $\dot{f}_{\rm 0}$ to get the effects coming from mass transfer and tidal interaction. Following that the interesting analysis related to matter property can be done then.
\section{Measurement accuracy estimation of the post-Keplerian parameters}
In this section we will use Fisher information matrix to estimate the measurement accuracy of the post-Keplerian parameters
\begin{align}
\Gamma_{ij}=\left\langle\frac{\partial h}{\partial p_i}|\frac{\partial h}{\partial p_j}\right\rangle,
\end{align}
where $p_i$ represents the $i$-th parameter among the 9 parameters $(A,f_0,\dot{f}_0,\iota_0,\dot{\iota}_0,e_0,\dot{e}_0,\dot{\gamma}_0,\phi_0)$. Inversing the Fisher matrix we can estimate the measurement error as
\begin{align}
\Delta p_i=\sqrt{\Gamma^{-1}_{ii}}.
\end{align}
The inner product in the above equation is defined as
\begin{align}
\left\langle g|h\right\rangle=\frac{2}{S_n(f_0)}\int_0^Tg(t)h(t)dt,\label{eq1}
\end{align}
where $f_0$ corresponds to the post-Keplerian parameter and the adiabatic approximation has been used \cite{10.1093.mnrasl.slaa183}.

Regarding to the sensitivity of a space-based detector, we use the following approximation (Eq.~(13) of \cite{Robson_2019})
\begin{align}
S_n(f)&=\frac{10}{3L^2}\left(P_{\rm OMS}+2(1+\cos^2(f/f_*))\frac{P_{\rm acc}}{(2\pi f)^4}\right)\times\nonumber\\
&\left(1+\frac{6}{10}\left(\frac{f}{f_*}\right)^2\right),\\
f_*&=c/(2\pi L).
\end{align}
For LISA \cite{Robson_2019} we have
\begin{align}
P_{\rm OMS}&=(1.5\times10^{-11}{\rm m})^2{\rm Hz}^{-1},\\
P_{\rm acc}&=(3\times10^{-15}{\rm ms}^{-2})^2\left(1+\left(\frac{4\times10^{-4}{\rm Hz}}{f}\right)^2\right){\rm Hz}^{-1},\\
L&=2.5\times10^9{\rm m}.
\end{align}
For Taiji \cite{PhysRevD.102.024089} we have
\begin{align}
P_{\rm OMS}&=(8\times10^{-12}{\rm m})^2{\rm Hz}^{-1},\\
P_{\rm acc}&=(3\times10^{-15}{\rm ms}^{-2})^2\left(1+\left(\frac{4\times10^{-4}{\rm Hz}}{f}\right)^2\right){\rm Hz}^{-1},\\
L&=3\times10^9{\rm m}.
\end{align}
For Tianqin we have \cite{luo2016tianqin}
\begin{align}
P_{\rm OMS}&=(1\times10^{-12}{\rm m})^2{\rm Hz}^{-1},\\
P_{\rm acc}&=(1\times10^{-15}{\rm ms}^{-2})^2\left(1+\left(\frac{1\times10^{-4}{\rm Hz}}{f}\right)^2\right){\rm Hz}^{-1},\\
L&=\sqrt{3}\times10^8{\rm m}.
\end{align}
\begin{figure}
\begin{tabular}{c}
\includegraphics[width=0.45\textwidth]{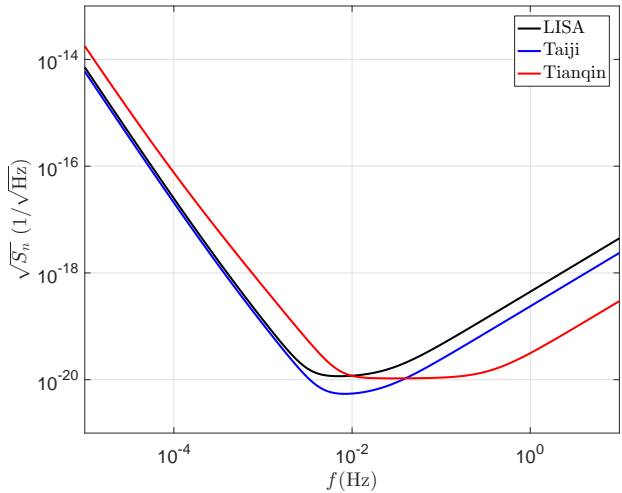}
\end{tabular}
\caption{Approximated sensitivity curves respectively for LISA \cite{Robson_2019}, Taiji \cite{PhysRevD.102.024089} and Tianqin \cite{luo2016tianqin} used in the current work. Note that the designed sensitivity of these three detectors is still changing because the final decision has not been reached yet. But such detail does not affect our analysis done in the current paper.}\label{fig1}
\end{figure}

Due to the relation (\ref{eq1}), the measurement accuracy for different detector is similar only upto the factor $1/S_n(f_0)$. Since this fact we calculate the measurement accuracy for LISA first and deduce the accuracy for other detectors accordingly in the following. As \cite{PhysRevD.102.063021}, we take the observation time as $T=5$ years.

\begin{table}[htb]
\caption{\label{tab1} Typical measurement accuracy of post-Keplerian parameters. The second column represents the assumed values taken by the parameters themselves. The rest three columns correspond to the measurement accuracy detected by LISA, Taiji and Tianqin respectively. Here four sets of parameters are investigated. Each set corresponds to one group listed below.}
\begin{center}
\begin{ruledtabular}
\begin{tabular}{c|c|c|c|c}
 & value & LISA & Taiji & Tianqin\\
\hline
$A$ & $10^{-20}$ & $9.5\times10^{-23}$ & $7.9\times10^{-23}$& $4.3\times10^{-22}$\\
$f_0$ & $10^{-3}$ & $3.1\times10^{-12}$ & $2.6\times10^{-12}$& $1.4\times10^{-11}$\\
$\dot{f}_0$ & $10^{-21}$ & $1.3\times10^{-20}$ & $1.1\times10^{-20}$ & $5.9\times10^{-20}$\\
$\iota_0$ & $\frac{\pi}{4}$ & $1.4\times10^{-2}$ & $1.2\times10^{-2}$& $6.4\times10^{-2}$\\
$\dot{\iota}_0$ & $0$ & $1.0\times10^{-11}$ & $8.7\times10^{-12}$& $4.7\times10^{-11}$\\
$e_0$ & $0.1$ & $2.6\times10^{-4}$ & $2.2\times10^{-4}$ & $1.2\times10^{-3}$\\
$\dot{e}_0$ & $10^{-18}$ & $2.9\times10^{-12}$ & $2.4\times10^{-12}$ & $1.3\times10^{-11}$\\
$\dot{\gamma}_0$ & $10^{-7}$ & $1.5\times10^{-11}$ & $1.2\times10^{-11}$& $6.7\times10^{-11}$\\
$\phi_0$ & $0$ & $4.6\times10^{-4}$ & $3.8\times10^{-4}$& $2.1\times10^{-3}$\\
\hline
$A$ & $10^{-20}$ & $9.4\times10^{-24}$ & $4.4\times10^{-24}$& $9.4\times10^{-24}$\\
$f_0$ & $10^{-2}$ & $3.1\times10^{-13}$ & $1.4\times10^{-13}$& $3.1\times10^{-13}$\\
$\dot{f}_0$ & $10^{-18}$ & $1.3\times10^{-21}$ & $6.1\times10^{-22}$ & $1.3\times10^{-21}$\\
$\iota_0$ & $\frac{\pi}{4}$ & $1.4\times10^{-3}$ & $6.6\times10^{-4}$& $1.4\times10^{-3}$\\
$\dot{\iota}_0$ & $0$ & $1.0\times10^{-12}$ & $4.8\times10^{-13}$& $1.0\times10^{-12}$\\
$e_0$ & $0.1$ & $2.6\times10^{-4}$ & $1.2\times10^{-5}$ & $2.6\times10^{-4}$\\
$\dot{e}_0$ & $10^{-16}$ & $2.9\times10^{-13}$ & $1.3\times10^{-13}$ & $2.9\times10^{-13}$\\
$\dot{\gamma}_0$ & $10^{-6}$ & $1.5\times10^{-12}$ & $6.8\times10^{-13}$& $1.5\times10^{-12}$\\
$\phi_0$ & $0$ & $4.6\times10^{-5}$ & $2.1\times10^{-5}$& $4.6\times10^{-5}$\\
\hline
$A$ & $10^{-20}$ & $1.4\times10^{-23}$ & $1.1\times10^{-23}$& $6.1\times10^{-23}$\\
$f_0$ & $10^{-3}$ & $8.0\times10^{-13}$ & $6.6\times10^{-13}$& $3.6\times10^{-12}$\\
$\dot{f}_0$ & $10^{-21}$ & $4.7\times10^{-21}$ & $3.9\times10^{-21}$ & $2.1\times10^{-20}$\\
$\iota_0$ & $\frac{\pi}{4}$ & $2.2\times10^{-3}$ & $1.8\times10^{-3}$& $9.7\times10^{-3}$\\
$\dot{\iota}_0$ & $0$ & $9.2\times10^{-12}$ & $7.7\times10^{-12}$& $4.1\times10^{-11}$\\
$e_0$ & $0.6$ & $1.8\times10^{-4}$ & $1.5\times10^{-4}$ & $8.1\times10^{-4}$\\
$\dot{e}_0$ & $10^{-18}$ & $2.0\times10^{-12}$ & $1.7\times10^{-12}$ & $8.9\times10^{-12}$\\
$\dot{\gamma}_0$ & $10^{-7}$ & $3.5\times10^{-12}$ & $2.9\times10^{-12}$& $1.6\times10^{-11}$\\
$\phi_0$ & $0$ & $1.7\times10^{-4}$ & $1.4\times10^{-4}$& $7.4\times10^{-4}$\\
\hline
$A$ & $10^{-20}$ & $9.5\times10^{-23}$ & $7.9\times10^{-23}$& $4.3\times10^{-22}$\\
$f_0$ & $10^{-3}$ & $3.1\times10^{-12}$ & $2.6\times10^{-12}$& $1.4\times10^{-11}$\\
$\dot{f}_0$ & $10^{-18}$ & $1.3\times10^{-20}$ & $1.1\times10^{-20}$ & $5.9\times10^{-20}$\\
$\iota_0$ & $\frac{\pi}{4}$ & $1.4\times10^{-2}$ & $1.2\times10^{-2}$& $6.4\times10^{-2}$\\
$\dot{\iota}_0$ & $0$ & $1.0\times10^{-11}$ & $8.7\times10^{-12}$& $4.7\times10^{-11}$\\
$e_0$ & $0.1$ & $2.7\times10^{-4}$ & $2.2\times10^{-4}$ & $1.2\times10^{-3}$\\
$\dot{e}_0$ & $10^{-16}$ & $2.9\times10^{-12}$ & $2.4\times10^{-12}$ & $1.3\times10^{-11}$\\
$\dot{\gamma}_0$ & $10^{-6}$ & $1.5\times10^{-11}$ & $1.2\times10^{-11}$& $6.7\times10^{-11}$\\
$\phi_0$ & $0$ & $4.6\times10^{-4}$ & $3.8\times10^{-4}$& $2.1\times10^{-3}$\\
 \end{tabular}
 \end{ruledtabular}
 \end{center}
\end{table}

\begin{figure*}
\begin{tabular}{c}
\includegraphics[width=\textwidth]{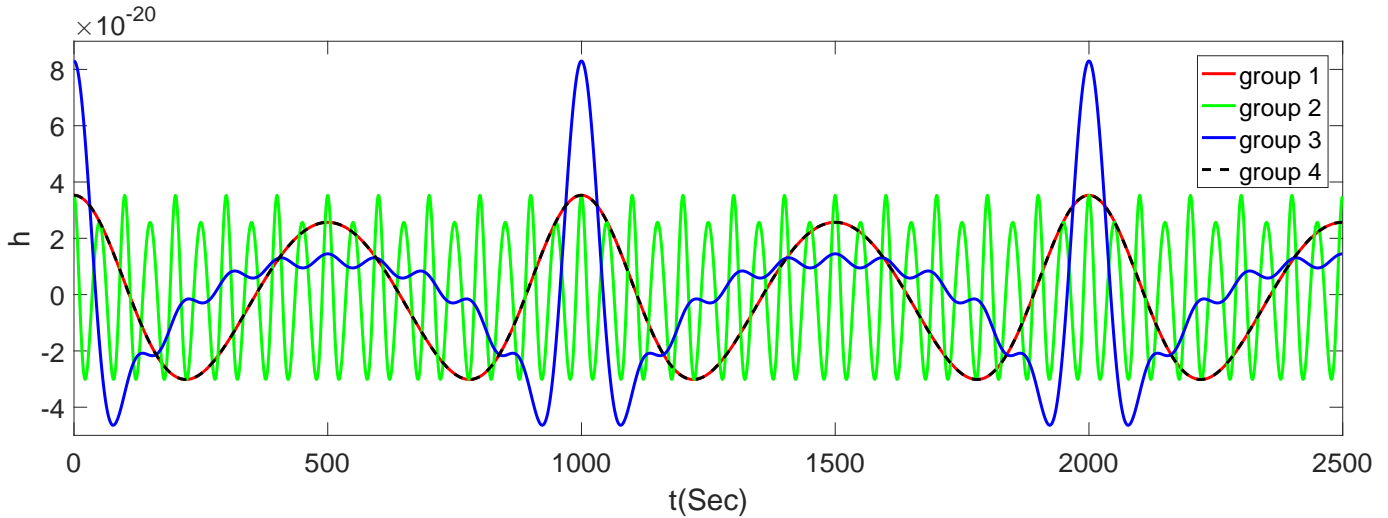}
\end{tabular}
\caption{Typical post-Keplerian waveforms correspond to the parameters shown in Table.~\ref{tab1}. Group 2 admits higher frequency than others. Group 3 admits higher eccentricity than others. Consequently the waveform of group 3 shows clear higher tone behavior. Compared to group 1, group 4 admits higher $\dot{e}_0$ and $\dot{\gamma}_0$. But these parameters difference only affect long term behavior. The short term waveform shown here can not indicate such difference.}\label{fig2}
\end{figure*}

We list the typical measurement accuracies of post-Keplerian parameters in Table.~\ref{tab1}. Four typical post-Keplerian parameters are investigated. Each set corresponds to a typical binary system. The corresponding waveforms of these four groups parameters are shown in Fig.~\ref{fig2}.

The first group corresponds to a typical binary white dwarf with pure GR effect. We can see all three detectors including LISA, Taiji and Tianqin can accurately measure almost parameters except $\dot{f}_0$ and $\dot{e}_0$. Here when we say `accurately' we mean the relative error is less than one percent. Regarding to $\phi_0$, it's value takes order 0.1 as the angle $\iota_0$ although the assumed value is 0 here. Regarding to $\dot{\iota}_0$, if the effect like (\ref{eq14}) comes in, its value will be order $10^{-8}$. Relative to these values, the measurement error bar is small.

The well estimated $\dot{\gamma}_0$ can be used to judge whether the matter effect is strong or not based on (\ref{eq6}). If the matter effect is ignorable, the well estimated $\dot{\gamma}_0$ can be used to constrain gravity theory beyond GR based on (\ref{eq16}). Note that $\frac{\Delta \dot{\gamma}_0}{\dot{\gamma}_0}\lesssim10^{-4}$, we can constrain
\begin{align}
&\left|\frac{\mathcal{P}}{\mathcal{G}}-1\right|\lesssim10^{-4},\\
&\left|2\gamma-\beta-1\right|\lesssim10^{-4}.
\end{align}
If we can further make sure the source is two black holes, we can get
\begin{align}
&\left|\mathcal{G}-1\right|\lesssim10^{-4},\\
&\left|\xi\right|\lesssim10^{-4},\\
&\left|\omega_{\rm BD}\right|\gtrsim10^{4}.
\end{align}

If everything indicates that only GR dominates, we can plug $\mathcal{M}$ and $D_L$ into the the Eqs.~(\ref{eq5})-(\ref{eq15}) and do the parameters estimation again. Then the chirp mass $\mathcal{M}$ can be well estimated \cite{PhysRevLett.100.041102,10.1093.mnrasl.slaa183}. With the information of $\dot{\gamma}_0$, $e_0$ and $f_0$, the Eq.~(\ref{eq15}) gives us the information of total mass $M$. Combining $M$ and $\mathcal{M}$ we can deduce mass ratio. Consequently the individual mass can be well estimated. And more we can use detail waveform model according to specific gravity theory and plug in component masses at this stage. Then the involved parameters may be constrained more strictly.

If the matter of the compact stars is very stiff, the two components may be very close. Possibly the frequency $f_0$ may be as high as $10^{-2}$Hz. We investigate such case in the second group of Table.~\ref{tab1}. This set of parameters also corresponds to stellar origin binary black hole or binary neutron star. If the mass transfer and the tidal interaction between two white dwarfs are very strong, the binary's frequency may also fall in this group. Firstly we can see LISA and Tianqin get roughly the same measurement accuracy. This is because the sensitivity for LISA and Tianqin roughly equals at $10^{-2}$Hz. Compared to the first group, the measurement accuracy improves roughly one order. This is due to that the detector sensitivity at $10^{-2}$Hz is about 10 times better than that at $10^{-3}$Hz. Consequently the afore mentioned parameter constrain and gravity theory constrain may be improved one order based on this kind of sources.

Compared to the first group, now we consider more eccentric binary with $e_0=0.6$ in the third group. Along with the increasing of the eccentricity, the waveform becomes more complicated. As shown in Fig.~\ref{fig2}, higher tones appear clearly \cite{Han_2017,Liu_2022}. As expected, higher eccentricity improves much the measurement accuracy \cite{PhysRevD.92.044034,PhysRevD.96.084046,PhysRevD.100.124003,PhysRevD.100.124032,PhysRevLett.129.191102,PhysRevD.107.043539}. Compared to the first group, the measurement accuracy improves one order. This means more eccentric binary even facilitate more the study of the gravity theory and the nuclear matter property.

In the fourth group we consider higher $\dot{f}_0$, $\dot{e}_0$ and $\dot{\gamma}_0$ compared to the first group. This may happen if the tidal interaction and the mass transfer between the two binary components take effect. We find that the measurement accuracy is roughly the same to the ones got in the first group. This result indicates only $A$, $f_0$ and $e_0$ affect the measurement accuracy strongly. At the same time we can conclude that the current planning space-based gravitational wave detectors can determine the post-Keplerian parameters as well as that shown in the first group of Table.~\ref{tab1}. If the signal is stronger (larger $A$), and/or the frequency is higher (larger $f_0$), and/or the orbit is more eccentric (larger $e_0$), we can get even better parameter estimation. Consequently we can constrain gravity theory tight and give deep insight to the nuclear matter.

\section{Conclusion and discussion}
Compact object binary provides ones a valuable laboratory for fundamental physics and astronomy. Pulsar timing detection has made a good use of binary pulsars, Existing gravitational wave detection has made a good use of coalescence of binary black hole, binary neutron star and neutron star-black hole binary. In the near future, the space-based gravitational wave detectors will facilitate us to make early inspiraling compact object binaries also become good laboratories. To do so, feasible waveform model is important.

In the current work, we propose a post-Keplerian waveform model for early inspiraling compact object binaries. Such a waveform model takes all kinds of gravity theories and all kinds of matter dynamics involved in the binary into consideration. This post-Keplerian waveform model can be looked as a unification of all previous existing waveform models for early inspiraling compact object binaries.

The post-Keplerian expansion parameters of the binary orbit are the waveform model parameters instead of binary component mass and other matter parameters. Based on this waveform model we can constrain gravity theory without the assumption of specific gravity theory. This trick is quite similar to that used in pulsar timing experiments. Based on the post-Keplerian expansion, we have constructed a ready-to-use unified waveform model of binary compact object for space-based detectors.

We have also checked the parameter estimation accuracy for the post-Keplerian parameters. Based on the current planning space-based gravitational wave detectors including LISA, Taiji and Tianqin, we find that the post-Keplerian parameters can be determined accurately. Consequently gravitational interaction dominating or matter dynamics dominating can be well distinguished. Following that the model independent constrain on gravity theory can be well done for gravitational dominated binary. And the nuclear matter properties can be well studied for matter dynamics dominated binary. Detail parameter estimation accuracy has been shown in Table.~\ref{tab1}. Using our waveform model, ones can not only catch the GW signal much easier but also distinguish different physical situations and even constraint different gravity theories based on LISA, Taiji or Tianqin detectors. When specific gravity theory or matter dynamics is determined, more specific waveform model can be used to fine tune the involved parameters.

Phenomenologically the effects of spin precession and orbital eccentricity will result in similar waveforms. Like event GW190521, it is hard to distinguish spin precession and orbital eccentricity for ground-based detectors \cite{2023MNRAS.519.5352R}. In the GW190521 case, we are limited by the small number of observed GW cycles. For the space-based detectors we considered in the current work, this will not be an issue. This is because the observation time is 5 years and many wave cycles will be accumulated.

In conclusion, our post-Keplerian waveform model provides a good toolkit for the future space-based gravitational wave detectors to study early inspiraling compact object binaries.
\section*{Acknowledgments}
This work was supported in part by the National Key Research and Development Program of China Grant
No. 2021YFC2203001 and the CAS Project for Young Scientists in Basic Research YSBR-006.

\appendix
\section{Waveform derivative respect to post-Keplerian parameters}
Based on the setting in the main text, the derivative respect to the $i$-th parameter can be written as
\begin{align}
&\frac{\partial h}{\partial p_i}=\frac{\sqrt{3}}{2}\sum_{n=1}^\infty[F_+\frac{\partial h_n^+}{\partial p_i}+F_\times \frac{\partial h_n^\times}{\partial p_i}]
\end{align}

Consequently we have
\begin{align}
&\int_0^T\frac{\partial h}{\partial p_i}\frac{\partial h}{\partial p_j}dt\nonumber\\
=&\frac{3}{4}\sum_{n,m=0}^\infty\int_0^T[F_+\frac{\partial h_n^+}{\partial p_i}+F_\times \frac{\partial h_n^\times}{\partial p_i}][F_+\frac{\partial h_m^+}{\partial p_j}+F_\times \frac{\partial h_m^\times}{\partial p_j}]dt\nonumber\\
=&\frac{3}{4}\sum_{n,m=0}^\infty[F^2_+\int_0^T\frac{\partial h_n^+}{\partial p_i}\frac{\partial h_m^+}{\partial p_j}dt+F^2_\times\int_0^T\frac{\partial h_n^\times}{\partial p_i}\frac{\partial h_m^\times}{\partial p_j}dt\nonumber\\
&+F_+F_\times\int_0^T(\frac{\partial h_n^+}{\partial p_i}\frac{\partial h_m^\times}{\partial p_j}+\frac{\partial h_n^\times}{\partial p_i}\frac{\partial h_m^+}{\partial p_j})dt]
\end{align}

Derivative respect to $A$
\begin{align}
&\frac{\partial h_n^+}{\partial A}=\{(1+\cos^2\iota)a_n(e)\cos[n\phi(t)]\cos(2\gamma)\nonumber\\
&-(1+\cos^2\iota)b_n(e)\sin[n\phi(t)]\sin(2\gamma)\nonumber\\
&+(\sin^2\iota) c_n(e)\cos[n\phi(t)]\},\\
&\frac{\partial h_n^\times}{\partial A}=-2\cos\iota\{b_n(e)\sin[n\phi(t)]\cos(2\gamma)\nonumber\\
&+a_n(e)\cos[n\phi(t)]\sin(2\gamma)\}.
\end{align}
Derivative respect to $\phi_0$
\begin{align}
&\frac{\partial h_n^+}{\partial \phi_0}=-A\{(1+\cos^2\iota)a_n(e)n\sin[n\phi(t)]\cos(2\gamma)\nonumber\\
&+(1+\cos^2\iota)b_n(e)n\cos(n\phi(t))\sin(2\gamma)]\nonumber\\
&+(\sin^2\iota) c_n(e)n\sin[n\phi(t)]\},\\
&\frac{\partial h_n^\times}{\partial \phi_0}=-2A\cos\iota\{b_n(e)n\cos[n\phi(t)]\cos(2\gamma)\nonumber\\
&-a_n(e)n\sin[n\phi(t)]\sin(2\gamma)\}.
\end{align}
Due to $\phi(t)$ dependence we have $\frac{\partial h}{\partial f_0}=2\pi t\frac{\partial h}{\partial \phi_0}$ and $\frac{\partial h}{\partial \dot{f}_0}=2\pi t^2\frac{\partial h}{\partial \phi_0}$. Derivative respect to $\iota_0$
\begin{align}
&\frac{\partial h_n^+}{\partial \iota_0}=-A\{a_n(e)\sin(2\iota)\cos[n\phi(t)]\cos(2\gamma)\nonumber\\
&-b_n(e)\sin(2\iota)\sin[n\phi(t)]\sin(2\gamma)\nonumber\\
&-c_n(e)\sin(2\iota)\cos[n\phi(t)]\},\\
&\frac{\partial h_n^\times}{\partial \iota_0}=2A\sin\iota\{a_n(e)\cos[n\phi(t)]\sin(2\gamma)\nonumber\\
&+b_n(e)\sin[n\phi(t)]\cos(2\gamma)\}.
\end{align}
Due to $\iota$ dependence we have $\frac{\partial h}{\partial \dot{\iota}_0}=t\frac{\partial h}{\partial \iota_0}$. Derivative respect to $e_0$
\begin{align}
&\frac{\partial h_n^+}{\partial e_0}=A\{(1+\cos^2\iota)a'_n(e)\cos[n\phi(t)]\cos(2\gamma)\nonumber\\
&-(1+\cos^2\iota)b'_n(e)\sin(n\phi(t))\sin(2\gamma)]\nonumber\\
&+(\sin^2\iota) c'_n(e)\cos[n\phi(t)]\},\\
&\frac{\partial h_n^\times}{\partial e_0}=-2A\cos\iota\{b'_n(e)\sin[n\phi(t)]\cos(2\gamma)\nonumber\\
&+a'_n(e)\cos[n\phi(t)]\sin(2\gamma)\},\\
&a'_n(e)=\frac{n^2}{2}[J_{n-3}(ne)-2eJ_{n-2}(ne)\nonumber\\
&-(\frac{2}{n}+1)J_{n-1}(ne)+4eJ_{n}(ne)\nonumber\\
&+(\frac{2}{n}-1)J_{n+1}(ne)-2eJ_{n+2}+J_{n+3}],\\
&b'_n(e)=-\frac{2ne}{\sqrt{1-e^2}}[J_{n-2}(ne)-2J_n(ne)+J_{n+2}(ne)]\nonumber\\
&+\frac{n^2\sqrt{1-e^2}}{2}[J_{n-3}(ne)-3J_{n-1}(ne)\nonumber\\
&+3J_{n+1}(ne)-J_{n+3}(ne)],\\
&c'_n(e)=\frac{n}{2}[J_{n-1}(ne)-J_{n+1}(ne)].
\end{align}
Due to $e$ dependence we have $\frac{\partial h}{\partial \dot{e}_0}=t\frac{\partial h}{\partial e_0}$. Derivative respect to $\dot{\gamma}_0$
\begin{align}
&\frac{\partial h_n^+}{\partial \dot{\gamma}_0}=-2At(1+\cos^2\iota)\{a_n(e)\cos[n\phi(t)]\sin(2\gamma)\nonumber\\
&+b_n(e)\sin[n\phi(t)]\cos(2\gamma)\},\\
&\frac{\partial h_n^\times}{\partial \dot{\gamma}_0}=-4A\cos\iota\{a_n(e)\cos[n\phi(t)]\cos(2\gamma)\nonumber\\
&-b_n(e)\sin[n\phi(t)]\sin(2\gamma)\}.
\end{align}

\section{Calculation of the inner product}
In order to calculate the inner product (\ref{eq1}), we need calculate the integrate $\int_0^Tg(t)h(t)dt$. Here functions $g(t)$ and $h$ take form
\begin{align}
g(t)=g(\phi(t);\vec{\xi}(t))
\end{align}
$\vec{\xi}$ are adiabatic parameters set including $(\iota,e,\gamma)$ which slowly change respect to time $t$. Consequently we have
\begin{align}
&g(t)\approx g(\phi(t);\vec{\xi}_0)+\frac{\partial g}{\partial\vec{\xi}}\cdot\dot{\vec{\xi}}\ t,\\
&\int_0^Tg(t)h(t)dt\approx\int_0^Tg(\phi(t);\vec{\xi}_0)h(\phi(t);\vec{\xi}_0)dt+\nonumber\\
&\dot{\vec{\xi}}\cdot\int_0^Tt\left[g(\phi(t);\vec{\xi}_0)\frac{\partial h}{\partial\vec{\xi}}+h(\phi(t);\vec{\xi}_0)\frac{\partial g}{\partial\vec{\xi}}\right]dt,\\
&\int_0^Tt\left[g(\phi(t);\vec{\xi}_0)\frac{\partial h}{\partial\vec{\xi}}+h(\phi(t);\vec{\xi}_0)\frac{\partial g}{\partial\vec{\xi}}\right]dt\approx\nonumber\\
&\frac{T}{2}\int_0^T\left[g(\phi(t);\vec{\xi}_0)\frac{\partial h}{\partial\vec{\xi}}+h(\phi(t);\vec{\xi}_0)\frac{\partial g}{\partial\vec{\xi}}\right]dt.
\end{align}
Related to the Fisher information matrix calculation we will met following form $\int_0^Tt^2g(\phi(t);\vec{\xi}_0)h(\phi(t);\vec{\xi}_0)dt$. We approximate it as
\begin{align}
&\int_0^Tt^2g(\phi(t);\vec{\xi}_0)h(\phi(t);\vec{\xi}_0)dt\approx\nonumber\\
&\frac{T^2}{3}\int_0^Tg(\phi(t);\vec{\xi}_0)h(\phi(t);\vec{\xi}_0)dt.
\end{align}

In addition we have relations
\begin{align}
&\int_0^T\sin(n\phi)\sin(m\phi)dt\nonumber\\
&\approx\int_0^\Phi\sin(n\phi)\sin(m\phi)\frac{d\phi}{2\pi f_0}\nonumber\\
&\approx\delta_{nm}\frac{\Phi}{4\pi f_0}=\frac{T}{2}\delta_{nm},\\
&\int_0^T\cos(n\phi)\cos(m\phi)dt\approx\frac{T}{2}\delta_{nm}\\
&\int_0^T\cos(n\phi)\sin(m\phi)dt\approx0,
\end{align}
which will be used in the Fisher information matrix calculation.

According to the above approximation we have
\begin{align}
&\int_0^T\frac{\partial h_n^+}{\partial A}(\phi(t);\vec{\xi}_0)\frac{\partial h_m^+}{\partial A}(\phi(t);\vec{\xi}_0)dt\approx\nonumber\\
&\int_0^T\{(1+\cos^2\iota_0)a_n(e_0)\cos[n\phi(t)]\cos(2\gamma_0)\nonumber\\
&-(1+\cos^2\iota_0)b_n(e_0)\sin[n\phi(t)]\sin(2\gamma_0)\nonumber\\
&+(\sin^2\iota_0) c_n(e_0)\cos[n\phi(t)]\}\times\nonumber\\
&\{(1+\cos^2\iota_0)a_m(e_0)\cos[m\phi(t)]\cos(2\gamma_0)\nonumber\\
&-(1+\cos^2\iota_0)b_m(e_0)\sin[m\phi(t)]\sin(2\gamma_0)\nonumber\\
&+(\sin^2\iota_0) c_m(e_0)\cos[m\phi(t)]\}dt\\
&\approx\frac{T\delta_{nm}}{2}\{[(1+\cos^2\iota_0)a_n(e_0)\cos(2\gamma_0)+(\sin^2\iota_0) c_n(e_0)]^2\nonumber\\
&+(1+\cos^2\iota_0)^2b^2_n(e_0)\sin^2(2\gamma_0)\},\\
&\int_0^Tt\frac{\partial h_n^+}{\partial A}(\phi(t);\vec{\xi}_0)\frac{\partial}{\partial \iota}\frac{\partial h_m^+}{\partial A}(\phi(t);\vec{\xi}_0)dt\approx\nonumber\\
&\frac{T}{2}\int_0^T\{(1+\cos^2\iota_0)a_n(e_0)\cos[n\phi(t)]\cos(2\gamma_0)\nonumber\\
&-(1+\cos^2\iota_0)b_n(e_0)\sin[n\phi(t)]\sin(2\gamma_0)\nonumber\\
&+(\sin^2\iota_0) c_n(e_0)\cos[n\phi(t)]\}\times\nonumber\\
&\{-\sin(2\iota_0)a_m(e_0)\cos[m\phi(t)]\cos(2\gamma_0)\nonumber\\
&+\sin(2\iota_0)b_m(e_0)\sin[m\phi(t)]\sin(2\gamma_0)\nonumber\\
&+\sin(2\iota_0) c_m(e_0)\cos[m\phi(t)]\}dt\\
&\approx\frac{T^2\delta_{nm}}{4}\{\sin(2\iota_0)[ c_n(e_0)-a_n(e_0)\cos(2\gamma_0)]\cdot\nonumber\\
&[(1+\cos^2\iota_0)a_n(e_0)\cos(2\gamma_0)+(\sin^2\iota_0) c_n(e_0)]\nonumber\\
&-(1+\cos^2\iota_0)\sin(2\iota_0)b^2_n(e_0)\sin^2(2\gamma_0)\},\\
&\int_0^Tt\frac{\partial h_n^+}{\partial A}(\phi(t);\vec{\xi}_0)\frac{\partial}{\partial e}\frac{\partial h_m^+}{\partial A}(\phi(t);\vec{\xi}_0)dt\approx\nonumber\\
&\frac{T}{2}\int_0^T\{(1+\cos^2\iota_0)a_n(e_0)\cos[n\phi(t)]\cos(2\gamma_0)\nonumber\\
&-(1+\cos^2\iota_0)b_n(e_0)\sin[n\phi(t)]\sin(2\gamma_0)\nonumber\\
&+(\sin^2\iota_0) c_n(e_0)\cos[n\phi(t)]\}\times\nonumber\\
&\{(1+\cos^2\iota_0)a'_n(e_0)\cos[m\phi(t)]\cos(2\gamma_0)\nonumber\\
&-(1+\cos^2\iota_0)b'_n(e_0)\sin[m\phi(t)]\sin(2\gamma_0)\nonumber\\
&+(\sin^2\iota_0) c'_n(e_0)\cos[m\phi(t)]\}dt\\
&\approx\frac{T^2\delta_{nm}}{4}\{[(1+\cos^2\iota_0)a_n(e_0)\cos(2\gamma_0)+(\sin^2\iota_0) c_n(e_0)]\cdot\nonumber\\
&[(1+\cos^2\iota_0)a'_n(e_0)\cos(2\gamma_0)+(\sin^2\iota_0) c'_n(e_0)]\nonumber\\
&+(1+\cos^2\iota_0)^2b_n(e_0)b'_n(e_0)\sin^2(2\gamma_0)\},\\
&\int_0^Tt\frac{\partial h_n^+}{\partial A}(\phi(t);\vec{\xi}_0)\frac{\partial}{\partial \gamma}\frac{\partial h_m^+}{\partial A}(\phi(t);\vec{\xi}_0)dt\approx\nonumber\\
&\frac{T}{2}\int_0^T\{(1+\cos^2\iota_0)a_n(e_0)\cos[n\phi(t)]\cos(2\gamma_0)\nonumber\\
&-(1+\cos^2\iota_0)b_n(e_0)\sin[n\phi(t)]\sin(2\gamma_0)\nonumber\\
&+(\sin^2\iota_0) c_n(e_0)\cos[n\phi(t)]\}\times\nonumber\\
&\{-2(1+\cos^2\iota_0)a_m(e_0)\cos[m\phi(t)]\sin(2\gamma_0)\nonumber\\
&-2(1+\cos^2\iota_0)b_m(e_0)\sin[m\phi(t)]\cos(2\gamma_0)\}dt\\
&\approx-\frac{T^2}{2}\delta_{nm}\{[(1+\cos^2\iota_0)a_n(e_0)\cos(2\gamma_0)+(\sin^2\iota_0) c_n(e_0)]\cdot\nonumber\\
&(1+\cos^2\iota_0)a_n(e_0)\sin(2\gamma_0)\nonumber\\
&+(1+\cos^2\iota_0)^2b^2_n(e_0)\sin(2\gamma_0)\cos(2\gamma_0)\},\\
&\int_0^T\frac{\partial h_n^+}{\partial A}\frac{\partial h_m^+}{\partial A}dt\approx\nonumber\\
&\frac{T\delta_{nm}}{2}\{[(1+\cos^2\iota_0)a_n(e_0)\cos(2\gamma_0)+(\sin^2\iota_0) c_n(e_0)]^2\nonumber\\
&+(1+\cos^2\iota_0)^2b^2_n(e_0)\sin^2(2\gamma_0)\}+\nonumber\\
&\frac{T^2\dot{\iota}_0\delta_{nm}}{2}\{\sin(2\iota_0)[ c_n(e_0)-a_n(e_0)\cos(2\gamma_0)]\cdot\nonumber\\
&[(1+\cos^2\iota_0)a_n(e_0)\cos(2\gamma_0)+(\sin^2\iota_0) c_n(e_0)]\nonumber\\
&-(1+\cos^2\iota_0)\sin(2\iota_0)b^2_n(e_0)\sin^2(2\gamma_0)\}+\nonumber\\
&\frac{T^2\dot{e}_0\delta_{nm}}{2}\{[(1+\cos^2\iota_0)a_n(e_0)\cos(2\gamma_0)+(\sin^2\iota_0) c_n(e_0)]\cdot\nonumber\\
&[(1+\cos^2\iota_0)a'_n(e_0)\cos(2\gamma_0)+(\sin^2\iota_0) c'_n(e_0)]\nonumber\\
&+(1+\cos^2\iota_0)^2b_n(e_0)b'_n(e_0)\sin^2(2\gamma_0)\}\nonumber\\
&-T^2\dot{\gamma}_0\delta_{nm}\{[(1+\cos^2\iota_0)a_n(e_0)\cos(2\gamma_0)+(\sin^2\iota_0) c_n(e_0)]\cdot\nonumber\\
&(1+\cos^2\iota_0)a_n(e_0)\sin(2\gamma_0)\nonumber\\
&+(1+\cos^2\iota_0)^2b^2_n(e_0)\sin(2\gamma_0)\cos(2\gamma_0)\}.\\
&\int_0^T\frac{\partial h_n^\times}{\partial A}(\phi(t);\vec{\xi}_0)\frac{\partial h_m^\times}{\partial A}(\phi(t);\vec{\xi}_0)dt\approx\nonumber\\
&\int_0^T4\cos^2\iota_0\{b_n(e_0)\sin[n\phi(t)]\cos(2\gamma_0)\nonumber\\
&+a_n(e_0)\cos[n\phi(t)]\sin(2\gamma_0)\}\times\nonumber\\
&\{b_m(e_0)\sin[m\phi(t)]\cos(2\gamma_0)\nonumber\\
&+a_m(e_0)\cos[m\phi(t)]\sin(2\gamma_0)\}dt\\
&\approx2T\cos^2\iota_0\delta_{nm}[b^2_n(e_0)\cos^2(2\gamma_0)+a^2_n(e_0)\sin^2(2\gamma_0)],\\
&\int_0^Tt\frac{\partial h_n^\times}{\partial A}(\phi(t);\vec{\xi}_0)\frac{\partial}{\partial \iota}\frac{\partial h_m^\times}{\partial A}(\phi(t);\vec{\xi}_0)dt\approx\nonumber\\
&\frac{T}{2}\int_0^T-4\sin\iota_0\cos\iota_0\{b_n(e_0)\sin[n\phi(t)]\cos(2\gamma_0)\nonumber\\
&+a_n(e_0)\cos[n\phi(t)]\sin(2\gamma_0)\}\times\nonumber\\
&\{b_m(e_0)\sin[m\phi(t)]\cos(2\gamma_0)\nonumber\\
&+a_m(e_0)\cos[m\phi(t)]\sin(2\gamma_0)\}dt\\
&\approx-T^2\sin\iota_0\cos\iota_0\delta_{nm}[b^2_n(e_0)\cos^2(2\gamma_0)\nonumber\\
&+a^2_n(e_0)\sin^2(2\gamma_0)],\\
&\int_0^Tt\frac{\partial h_n^\times}{\partial A}(\phi(t);\vec{\xi}_0)\frac{\partial}{\partial e}\frac{\partial h_m^\times}{\partial A}(\phi(t);\vec{\xi}_0)dt\approx\nonumber\\
&\frac{T}{2}\int_0^T4\cos^2\iota_0\{b_n(e_0)\sin[n\phi(t)]\cos(2\gamma_0)\nonumber\\
&+a_n(e_0)\cos[n\phi(t)]\sin(2\gamma_0)\}\times\nonumber\\
&\{b'_m(e_0)\sin[m\phi(t)]\cos(2\gamma_0)\nonumber\\
&+a'_m(e_0)\cos[m\phi(t)]\sin(2\gamma_0)\}dt\\
&\approx T^2\cos^2\iota_0\delta_{nm}[b_n(e_0)b'_n(e_0)\cos^2(2\gamma_0)\nonumber\\
&+a_n(e_0)a'_n(e_0)\sin^2(2\gamma_0)],\\
&\int_0^Tt\frac{\partial h_n^\times}{\partial A}(\phi(t);\vec{\xi}_0)\frac{\partial}{\partial \gamma}\frac{\partial h_m^\times}{\partial A}(\phi(t);\vec{\xi}_0)dt\approx\nonumber\\
&\frac{T}{2}\int_0^T4\cos^2\iota_0\{b_n(e_0)\sin[n\phi(t)]\cos(2\gamma_0)\nonumber\\
&+a_n(e_0)\cos[n\phi(t)]\sin(2\gamma_0)\}\times\nonumber\\
&\{-2b_m(e_0)\sin[m\phi(t)]\sin(2\gamma_0)\nonumber\\
&+2a_m(e_0)\cos[m\phi(t)]\cos(2\gamma_0)\}dt\\
&\approx2T^2\cos^2\iota_0\sin(2\gamma_0)\cos(2\gamma_0)\delta_{nm}[a^2_n(e_0)+b^2_n(e_0)],\\
&\int_0^T\frac{\partial h_n^\times}{\partial A}\frac{\partial h_m^\times}{\partial A}dt\approx\nonumber\\
&2T\cos^2\iota_0\delta_{nm}[b^2_n(e_0)\cos^2(2\gamma_0)+a^2_n(e_0)\sin^2(2\gamma_0)]\nonumber\\
&-2\dot{\iota}_0T^2\sin\iota_0\cos\iota_0\delta_{nm}[b^2_n(e_0)\cos^2(2\gamma_0)\nonumber\\
&+a^2_n(e_0)\sin^2(2\gamma_0)]\nonumber\\
&+2\dot{e}_0T^2\cos^2\iota_0\delta_{nm}[b_n(e_0)b'_n(e_0)\cos^2(2\gamma_0)\nonumber\\
&+a_n(e_0)a'_n(e_0)\sin^2(2\gamma_0)]\nonumber\\
&+4\dot{\gamma}_0T^2\cos^2\iota_0\sin(2\gamma_0)\cos(2\gamma_0)\delta_{nm}[a^2_n(e_0)+b^2_n(e_0)],\\
&\int_0^T\frac{\partial h_n^+}{\partial A}(\phi(t);\vec{\xi}_0)\frac{\partial h_m^\times}{\partial A}(\phi(t);\vec{\xi}_0)dt\approx\nonumber\\
&-\int_0^T\{(1+\cos^2\iota_0)a_n(e_0)\cos[n\phi(t)]\cos(2\gamma_0)\nonumber\\
&-(1+\cos^2\iota_0)b_n(e_0)\sin[n\phi(t)]\sin(2\gamma_0)\nonumber\\
&+(\sin^2\iota_0) c_n(e_0)\cos[n\phi(t)]\}\times\nonumber\\
&2\cos(\iota_0)\{b_m(e_0)\sin[m\phi(t)]\cos(2\gamma_0)\nonumber\\
&+a_m(e_0)\cos[m\phi(t)]\sin(2\gamma_0)\}dt\\
&\approx-T\delta_{nm}\cos(\iota_0)\{a_n(e_0)\sin(2\gamma_0)\times\nonumber\\
&[(1+\cos^2\iota_0)a_n(e_0)\cos(2\gamma_0)+(\sin^2\iota_0) c_n(e_0)]-\nonumber\\
&(1+\cos^2\iota_0)b^2_n(e_0)\sin(2\gamma_0)\cos(2\gamma_0)\},\\
&\int_0^Tt\frac{\partial}{\partial\iota}\frac{\partial h_n^+}{\partial A}(\phi(t);\vec{\xi}_0)\frac{\partial h_m^\times}{\partial A}(\phi(t);\vec{\xi}_0)dt\approx\nonumber\\
&-\frac{T}{2}\int_0^T\{-2\cos\iota_0\sin\iota_0a_n(e_0)\cos[n\phi(t)]\cos(2\gamma_0)\nonumber\\
&+2\cos\iota_0\sin\iota_0b_n(e_0)\sin[n\phi(t)]\sin(2\gamma_0)\nonumber\\
&+2\cos\iota_0\sin\iota_0 c_n(e_0)\cos[n\phi(t)]\}\times\nonumber\\
&2\cos(\iota_0)\{b_m(e_0)\sin[m\phi(t)]\cos(2\gamma_0)\nonumber\\
&+a_m(e_0)\cos[m\phi(t)]\sin(2\gamma_0)\}dt\\
&\approx-T^2\delta_{nm}\cos^2(\iota_0)\sin\iota_0\sin(2\gamma_0)\{a_n(e_0)\times\nonumber\\
&[c_n(e_0)-a_n(e_0)\cos(2\gamma_0)]+b^2_n(e_0)\cos(2\gamma_0)\},\\
&\int_0^Tt\frac{\partial h_n^+}{\partial A}(\phi(t);\vec{\xi}_0)\frac{\partial}{\partial\iota}\frac{\partial h_m^\times}{\partial A}(\phi(t);\vec{\xi}_0)dt\approx\nonumber\\
&T\int_0^T\{(1+\cos^2\iota_0)a_n(e_0)\cos[n\phi(t)]\cos(2\gamma_0)\nonumber\\
&-(1+\cos^2\iota_0)b_n(e_0)\sin[n\phi(t)]\sin(2\gamma_0)\nonumber\\
&+(\sin^2\iota_0) c_n(e_0)\cos[n\phi(t)]\}\times\nonumber\\
&\sin(\iota_0)\{b_m(e_0)\sin[m\phi(t)]\cos(2\gamma_0)\nonumber\\
&+a_m(e_0)\cos[m\phi(t)]\sin(2\gamma_0)\}dt\\
&\approx\frac{T^2\sin(\iota_0)\delta_{nm}}{2}\{[(1+\cos^2\iota_0)a_n(e_0)\cos(2\gamma_0)+\nonumber\\
&(\sin^2\iota_0) c_n(e_0)]a_n(e_0)\sin(2\gamma_0)-\nonumber\\
&(1+\cos^2\iota_0)b^2_n(e_0)\sin(2\gamma_0)\cos(2\gamma_0)\},\\
&\int_0^Tt\frac{\partial}{\partial e}\frac{\partial h_n^+}{\partial A}(\phi(t);\vec{\xi}_0)\frac{\partial h_m^\times}{\partial A}(\phi(t);\vec{\xi}_0)dt\approx\nonumber\\
&-\frac{T}{2}\int_0^T\{(1+\cos^2\iota_0)a'_n(e_0)\cos[n\phi(t)]\cos(2\gamma_0)\nonumber\\
&-(1+\cos^2\iota_0)b'_n(e_0)\sin[n\phi(t)]\sin(2\gamma_0)\nonumber\\
&+(\sin^2\iota_0) c'_n(e_0)\cos[n\phi(t)]\}\times\nonumber\\
&2\cos(\iota_0)\{b_m(e_0)\sin[m\phi(t)]\cos(2\gamma_0)\nonumber\\
&+a_m(e_0)\cos[m\phi(t)]\sin(2\gamma_0)\}dt\\
&\approx-\frac{T^2}{2}\delta_{nm}\cos(\iota_0)\{a_n(e_0)\sin(2\gamma_0)\times\nonumber\\
&[(1+\cos^2\iota_0)a'_n(e_0)\cos(2\gamma_0)+(\sin^2\iota_0) c'_n(e_0)]-\nonumber\\
&(1+\cos^2\iota_0)b_n(e_0)b'_n(e_0)\sin(2\gamma_0)\cos(2\gamma_0)\},\\
&\int_0^Tt\frac{\partial h_n^+}{\partial A}(\phi(t);\vec{\xi}_0)\frac{\partial}{\partial e}\frac{\partial h_m^\times}{\partial A}(\phi(t);\vec{\xi}_0)dt\approx\nonumber\\
&-\frac{T}{2}\int_0^T\{(1+\cos^2\iota_0)a_n(e_0)\cos[n\phi(t)]\cos(2\gamma_0)\nonumber\\
&-(1+\cos^2\iota_0)b_n(e_0)\sin[n\phi(t)]\sin(2\gamma_0)\nonumber\\
&+(\sin^2\iota_0) c_n(e_0)\cos[n\phi(t)]\}\times\nonumber\\
&2\cos(\iota_0)\{b'_m(e_0)\sin[m\phi(t)]\cos(2\gamma_0)\nonumber\\
&+a'_m(e_0)\cos[m\phi(t)]\sin(2\gamma_0)\}dt\\
&\approx-\frac{T^2}{2}\delta_{nm}\cos(\iota_0)\{a'_n(e_0)\sin(2\gamma_0)\times\nonumber\\
&[(1+\cos^2\iota_0)a_n(e_0)\cos(2\gamma_0)+(\sin^2\iota_0) c_n(e_0)]-\nonumber\\
&(1+\cos^2\iota_0)b_n(e_0)b'_n(e_0)\sin(2\gamma_0)\cos(2\gamma_0)\},\\
&\int_0^Tt\frac{\partial}{\partial\gamma}\frac{\partial h_n^+}{\partial A}(\phi(t);\vec{\xi}_0)\frac{\partial h_m^\times}{\partial A}(\phi(t);\vec{\xi}_0)dt\approx\nonumber\\
&T\int_0^T\{(1+\cos^2\iota_0)a_n(e_0)\cos[n\phi(t)]\sin(2\gamma_0)\nonumber\\
&+(1+\cos^2\iota_0)b_n(e_0)\sin[n\phi(t)]\cos(2\gamma_0)\}\times\nonumber\\
&2\cos(\iota_0)\{b_m(e_0)\sin[m\phi(t)]\cos(2\gamma_0)\nonumber\\
&+a_m(e_0)\cos[m\phi(t)]\sin(2\gamma_0)\}dt\\
&\approx T^2\delta_{nm}(1+\cos^2\iota_0)\cos(\iota_0)\times\nonumber\\
&[a^2_n(e_0)\sin^2(2\gamma_0)+b^2_n(e_0)\cos^2(2\gamma_0)],\\
&\int_0^Tt\frac{\partial h_n^+}{\partial A}(\phi(t);\vec{\xi}_0)\frac{\partial}{\partial\gamma}\frac{\partial h_m^\times}{\partial A}(\phi(t);\vec{\xi}_0)dt\approx\nonumber\\
&-\frac{T}{2}\int_0^T\{(1+\cos^2\iota_0)a_n(e_0)\cos[n\phi(t)]\cos(2\gamma_0)\nonumber\\
&-(1+\cos^2\iota_0)b_n(e_0)\sin[n\phi(t)]\sin(2\gamma_0)\nonumber\\
&+(\sin^2\iota_0) c_n(e_0)\cos[n\phi(t)]\}\times\nonumber\\
&2\cos(\iota_0)\{-2b_m(e_0)\sin[m\phi(t)]\sin(2\gamma_0)\nonumber\\
&+2a_m(e_0)\cos[m\phi(t)]\cos(2\gamma_0)\}dt\\
&\approx-T^2\delta_{nm}\cos(\iota_0)\{[(1+\cos^2\iota_0)a_n(e_0)\cos(2\gamma_0)+\nonumber\\
&(\sin^2\iota_0) c_n(e_0)]a_n(e_0)\cos(2\gamma_0)+\nonumber\\
&(1+\cos^2\iota_0)b^2_n(e_0)\sin^2(2\gamma_0)\},\\
&\int_0^T\frac{\partial h_n^+}{\partial A}\frac{\partial h_m^\times}{\partial A}dt\approx\nonumber\\
&-T\delta_{nm}\cos(\iota_0)\{a_n(e_0)\sin(2\gamma_0)\times\nonumber\\
&[(1+\cos^2\iota_0)a_n(e_0)\cos(2\gamma_0)+(\sin^2\iota_0) c_n(e_0)]-\nonumber\\
&(1+\cos^2\iota_0)b^2_n(e_0)\sin(2\gamma_0)\cos(2\gamma_0)\}\nonumber\\
&-T^2\dot{\iota}_0\delta_{nm}[\cos^2(\iota_0)\sin\iota_0\sin(2\gamma_0)\{a_n(e_0)\times\nonumber\\
&[c_n(e_0)-a_n(e_0)\cos(2\gamma_0)]+b^2_n(e_0)\cos(2\gamma_0)\}\\
&-\frac{\sin(\iota_0)}{2}\{[(1+\cos^2\iota_0)a_n(e_0)\cos(2\gamma_0)+\nonumber\\
&(\sin^2\iota_0) c_n(e_0)]a_n(e_0)\sin(2\gamma_0)-\nonumber\\
&(1+\cos^2\iota_0)b^2_n(e_0)\sin(2\gamma_0)\cos(2\gamma_0)\}]\nonumber\\
&-\frac{T^2}{2}\dot{e}_0\delta_{nm}\cos(\iota_0)\{a_n(e_0)\sin(2\gamma_0)\times\nonumber\\
&[(1+\cos^2\iota_0)a'_n(e_0)\cos(2\gamma_0)+(\sin^2\iota_0) c'_n(e_0)]-\nonumber\\
&(1+\cos^2\iota_0)b_n(e_0)b'_n(e_0)\sin(2\gamma_0)\cos(2\gamma_0)\nonumber\\
&+a'_n(e_0)\sin(2\gamma_0)\times\nonumber\\
&[(1+\cos^2\iota_0)a_n(e_0)\cos(2\gamma_0)+(\sin^2\iota_0) c_n(e_0)]-\nonumber\\
&(1+\cos^2\iota_0)b_n(e_0)b'_n(e_0)\sin(2\gamma_0)\cos(2\gamma_0)\}\nonumber\\
&+T^2\dot{\gamma}_0\delta_{nm}\cos(\iota_0)\{(1+\cos^2\iota_0)\times\nonumber\\
&[a^2_n(e_0)\sin^2(2\gamma_0)+b^2_n(e_0)\cos^2(2\gamma_0)]\\
&-\{[(1+\cos^2\iota_0)a_n(e_0)\cos(2\gamma_0)+\nonumber\\
&(\sin^2\iota_0) c_n(e_0)]a_n(e_0)\cos(2\gamma_0)+\nonumber\\
&(1+\cos^2\iota_0)b^2_n(e_0)\sin^2(2\gamma_0)\}\}.
\end{align}
The calculation of other elements of the Fisher information matrix is similar.

\section*{Data availability}
The data that support the plots within this paper and the other findings of this
study are available from the corresponding author upon request.

\bibliography{refs}

\begin{thebibliography}{69}
\expandafter\ifx\csname natexlab\endcsname\relax\def\natexlab#1{#1}\fi
\expandafter\ifx\csname bibnamefont\endcsname\relax
  \def\bibnamefont#1{#1}\fi
\expandafter\ifx\csname bibfnamefont\endcsname\relax
  \def\bibfnamefont#1{#1}\fi
\expandafter\ifx\csname citenamefont\endcsname\relax
  \def\citenamefont#1{#1}\fi
\expandafter\ifx\csname url\endcsname\relax
  \def\url#1{\texttt{#1}}\fi
\expandafter\ifx\csname urlprefix\endcsname\relax\def\urlprefix{URL }\fi
\providecommand{\bibinfo}[2]{#2}
\providecommand{\eprint}[2][]{\url{#2}}

\bibitem[{\citenamefont{Kramer et~al.}(2021)\citenamefont{Kramer, Stairs,
  Manchester, Wex, Deller, Coles, Ali, Burgay, Camilo, Cognard
  et~al.}}]{PhysRevX.11.041050}
\bibinfo{author}{\bibfnamefont{M.}~\bibnamefont{Kramer}},
  \bibinfo{author}{\bibfnamefont{I.~H.} \bibnamefont{Stairs}},
  \bibinfo{author}{\bibfnamefont{R.~N.} \bibnamefont{Manchester}},
  \bibinfo{author}{\bibfnamefont{N.}~\bibnamefont{Wex}},
  \bibinfo{author}{\bibfnamefont{A.~T.} \bibnamefont{Deller}},
  \bibinfo{author}{\bibfnamefont{W.~A.} \bibnamefont{Coles}},
  \bibinfo{author}{\bibfnamefont{M.}~\bibnamefont{Ali}},
  \bibinfo{author}{\bibfnamefont{M.}~\bibnamefont{Burgay}},
  \bibinfo{author}{\bibfnamefont{F.}~\bibnamefont{Camilo}},
  \bibinfo{author}{\bibfnamefont{I.}~\bibnamefont{Cognard}},
  \bibnamefont{et~al.}, \bibinfo{journal}{Phys. Rev. X}
  \textbf{\bibinfo{volume}{11}}, \bibinfo{pages}{041050}
  (\bibinfo{year}{2021}),
  \urlprefix\url{https://link.aps.org/doi/10.1103/PhysRevX.11.041050}.

\bibitem[{\citenamefont{Lorimer and Kramer}(2004)}]{2004Lorimer}
\bibinfo{author}{\bibfnamefont{D.~R.} \bibnamefont{Lorimer}} \bibnamefont{and}
  \bibinfo{author}{\bibfnamefont{M.}~\bibnamefont{Kramer}},
  \emph{\bibinfo{title}{{Handbook of Pulsar Astronomy, Vol. 4}}}
  (\bibinfo{publisher}{Cambridge University Press, Cambridge, England},
  \bibinfo{year}{2004}).

\bibitem[{\citenamefont{{The LIGO Scientific Collaboration}
  et~al.}(2021)\citenamefont{{The LIGO Scientific Collaboration}, {the Virgo
  Collaboration}, {the KAGRA Collaboration}, {Abbott}, {Abbott}, {Acernese},
  {Ackley}, {Adams}, {Adhikari}, {Adhikari} et~al.}}]{2021arXiv211103606T}
\bibinfo{author}{\bibnamefont{{The LIGO Scientific Collaboration}}},
  \bibinfo{author}{\bibnamefont{{the Virgo Collaboration}}},
  \bibinfo{author}{\bibnamefont{{the KAGRA Collaboration}}},
  \bibinfo{author}{\bibfnamefont{R.}~\bibnamefont{{Abbott}}},
  \bibinfo{author}{\bibfnamefont{T.~D.} \bibnamefont{{Abbott}}},
  \bibinfo{author}{\bibfnamefont{F.}~\bibnamefont{{Acernese}}},
  \bibinfo{author}{\bibfnamefont{K.}~\bibnamefont{{Ackley}}},
  \bibinfo{author}{\bibfnamefont{C.}~\bibnamefont{{Adams}}},
  \bibinfo{author}{\bibfnamefont{N.}~\bibnamefont{{Adhikari}}},
  \bibinfo{author}{\bibfnamefont{R.~X.} \bibnamefont{{Adhikari}}},
  \bibnamefont{et~al.}, \bibinfo{journal}{arXiv e-prints}
  \bibinfo{eid}{arXiv:2111.03606} (\bibinfo{year}{2021}), \eprint{2111.03606}.

\bibitem[{\citenamefont{Willems et~al.}(2008)\citenamefont{Willems, Vecchio,
  and Kalogera}}]{PhysRevLett.100.041102}
\bibinfo{author}{\bibfnamefont{B.}~\bibnamefont{Willems}},
  \bibinfo{author}{\bibfnamefont{A.}~\bibnamefont{Vecchio}}, \bibnamefont{and}
  \bibinfo{author}{\bibfnamefont{V.}~\bibnamefont{Kalogera}},
  \bibinfo{journal}{Phys. Rev. Lett.} \textbf{\bibinfo{volume}{100}},
  \bibinfo{pages}{041102} (\bibinfo{year}{2008}),
  \urlprefix\url{https://link.aps.org/doi/10.1103/PhysRevLett.100.041102}.

\bibitem[{\citenamefont{Huang et~al.}(2020)\citenamefont{Huang, Hu, Korol, Li,
  Liang, Lu, Wang, Yu, and Mei}}]{PhysRevD.102.063021}
\bibinfo{author}{\bibfnamefont{S.-J.} \bibnamefont{Huang}},
  \bibinfo{author}{\bibfnamefont{Y.-M.} \bibnamefont{Hu}},
  \bibinfo{author}{\bibfnamefont{V.}~\bibnamefont{Korol}},
  \bibinfo{author}{\bibfnamefont{P.-C.} \bibnamefont{Li}},
  \bibinfo{author}{\bibfnamefont{Z.-C.} \bibnamefont{Liang}},
  \bibinfo{author}{\bibfnamefont{Y.}~\bibnamefont{Lu}},
  \bibinfo{author}{\bibfnamefont{H.-T.} \bibnamefont{Wang}},
  \bibinfo{author}{\bibfnamefont{S.}~\bibnamefont{Yu}}, \bibnamefont{and}
  \bibinfo{author}{\bibfnamefont{J.}~\bibnamefont{Mei}},
  \bibinfo{journal}{Phys. Rev. D} \textbf{\bibinfo{volume}{102}},
  \bibinfo{pages}{063021} (\bibinfo{year}{2020}),
  \urlprefix\url{https://link.aps.org/doi/10.1103/PhysRevD.102.063021}.

\bibitem[{\citenamefont{{Nelemans} et~al.}(2001)\citenamefont{{Nelemans},
  {Yungelson}, {Portegies Zwart}, and {Verbunt}}}]{2001A&A...365..491N}
\bibinfo{author}{\bibfnamefont{G.}~\bibnamefont{{Nelemans}}},
  \bibinfo{author}{\bibfnamefont{L.~R.} \bibnamefont{{Yungelson}}},
  \bibinfo{author}{\bibfnamefont{S.~F.} \bibnamefont{{Portegies Zwart}}},
  \bibnamefont{and}
  \bibinfo{author}{\bibfnamefont{F.}~\bibnamefont{{Verbunt}}},
  \bibinfo{journal}{Astronomy \& Astrophysics} \textbf{\bibinfo{volume}{365}},
  \bibinfo{pages}{491} (\bibinfo{year}{2001}), \eprint{astro-ph/0010457}.

\bibitem[{\citenamefont{Seto}(2022)}]{PhysRevLett.128.041101}
\bibinfo{author}{\bibfnamefont{N.}~\bibnamefont{Seto}}, \bibinfo{journal}{Phys.
  Rev. Lett.} \textbf{\bibinfo{volume}{128}}, \bibinfo{pages}{041101}
  (\bibinfo{year}{2022}),
  \urlprefix\url{https://link.aps.org/doi/10.1103/PhysRevLett.128.041101}.

\bibitem[{\citenamefont{Cao and Han}(2017)}]{PhysRevD.96.044028_SEOBNRE}
\bibinfo{author}{\bibfnamefont{Z.}~\bibnamefont{Cao}} \bibnamefont{and}
  \bibinfo{author}{\bibfnamefont{W.-B.} \bibnamefont{Han}},
  \bibinfo{journal}{Phys. Rev. D} \textbf{\bibinfo{volume}{96}},
  \bibinfo{pages}{044028} (\bibinfo{year}{2017}),
  \urlprefix\url{https://link.aps.org/doi/10.1103/PhysRevD.96.044028}.

\bibitem[{\citenamefont{Liu et~al.}(2020)\citenamefont{Liu, Cao, and
  Shao}}]{PhysRevD.101.044049_validSEOBNRE}
\bibinfo{author}{\bibfnamefont{X.}~\bibnamefont{Liu}},
  \bibinfo{author}{\bibfnamefont{Z.}~\bibnamefont{Cao}}, \bibnamefont{and}
  \bibinfo{author}{\bibfnamefont{L.}~\bibnamefont{Shao}},
  \bibinfo{journal}{Phys. Rev. D} \textbf{\bibinfo{volume}{101}},
  \bibinfo{pages}{044049} (\bibinfo{year}{2020}),
  \urlprefix\url{https://link.aps.org/doi/10.1103/PhysRevD.101.044049}.

\bibitem[{\citenamefont{Liu et~al.}(2022)\citenamefont{Liu, Cao, and
  Zhu}}]{Liu_2022}
\bibinfo{author}{\bibfnamefont{X.}~\bibnamefont{Liu}},
  \bibinfo{author}{\bibfnamefont{Z.}~\bibnamefont{Cao}}, \bibnamefont{and}
  \bibinfo{author}{\bibfnamefont{Z.-H.} \bibnamefont{Zhu}},
  \bibinfo{journal}{Classical and Quantum Gravity}
  \textbf{\bibinfo{volume}{39}}, \bibinfo{pages}{035009}
  (\bibinfo{year}{2022}),
  \urlprefix\url{https://doi.org/10.1088/1361-6382/ac4119}.

\bibitem[{\citenamefont{{Takahashi} and {Seto}}(2002)}]{2002ApJ...575.1030T}
\bibinfo{author}{\bibfnamefont{R.}~\bibnamefont{{Takahashi}}} \bibnamefont{and}
  \bibinfo{author}{\bibfnamefont{N.}~\bibnamefont{{Seto}}},
  \bibinfo{journal}{The Astrophysical Journal} \textbf{\bibinfo{volume}{575}},
  \bibinfo{pages}{1030} (\bibinfo{year}{2002}), \eprint{astro-ph/0204487}.

\bibitem[{\citenamefont{Kremer et~al.}(2017)\citenamefont{Kremer, Breivik,
  Larson, and Kalogera}}]{Kremer_2017}
\bibinfo{author}{\bibfnamefont{K.}~\bibnamefont{Kremer}},
  \bibinfo{author}{\bibfnamefont{K.}~\bibnamefont{Breivik}},
  \bibinfo{author}{\bibfnamefont{S.~L.} \bibnamefont{Larson}},
  \bibnamefont{and} \bibinfo{author}{\bibfnamefont{V.}~\bibnamefont{Kalogera}},
  \bibinfo{journal}{The Astrophysical Journal} \textbf{\bibinfo{volume}{846}},
  \bibinfo{pages}{95} (\bibinfo{year}{2017}),
  \urlprefix\url{https://doi.org/10.3847/1538-4357/aa8557}.

\bibitem[{\citenamefont{Benacquista}(2011)}]{Benacquista_2011}
\bibinfo{author}{\bibfnamefont{M.~J.} \bibnamefont{Benacquista}},
  \bibinfo{journal}{The Astrophysical Journal} \textbf{\bibinfo{volume}{740}},
  \bibinfo{pages}{L54} (\bibinfo{year}{2011}),
  \urlprefix\url{https://doi.org/10.1088/2041-8205/740/2/l54}.

\bibitem[{\citenamefont{Wolz et~al.}(2020)\citenamefont{Wolz, Yagi, Anderson,
  and Taylor}}]{10.1093.mnrasl.slaa183}
\bibinfo{author}{\bibfnamefont{A.}~\bibnamefont{Wolz}},
  \bibinfo{author}{\bibfnamefont{K.}~\bibnamefont{Yagi}},
  \bibinfo{author}{\bibfnamefont{N.}~\bibnamefont{Anderson}}, \bibnamefont{and}
  \bibinfo{author}{\bibfnamefont{A.~J.} \bibnamefont{Taylor}},
  \bibinfo{journal}{Monthly Notices of the Royal Astronomical Society: Letters}
  \textbf{\bibinfo{volume}{500}}, \bibinfo{pages}{L52} (\bibinfo{year}{2020}),
  ISSN \bibinfo{issn}{1745-3925},
  \eprint{https://academic.oup.com/mnrasl/article-pdf/500/1/L52/34545852/slaa183.pdf},
  \urlprefix\url{https://doi.org/10.1093/mnrasl/slaa183}.

\bibitem[{\citenamefont{Damour and Taylor}(1992)}]{PhysRevD.45.1840}
\bibinfo{author}{\bibfnamefont{T.}~\bibnamefont{Damour}} \bibnamefont{and}
  \bibinfo{author}{\bibfnamefont{J.~H.} \bibnamefont{Taylor}},
  \bibinfo{journal}{Phys. Rev. D} \textbf{\bibinfo{volume}{45}},
  \bibinfo{pages}{1840} (\bibinfo{year}{1992}),
  \urlprefix\url{https://link.aps.org/doi/10.1103/PhysRevD.45.1840}.

\bibitem[{\citenamefont{Moore et~al.}(2016)\citenamefont{Moore, Favata, Arun,
  and Mishra}}]{PhysRevD.93.124061}
\bibinfo{author}{\bibfnamefont{B.}~\bibnamefont{Moore}},
  \bibinfo{author}{\bibfnamefont{M.}~\bibnamefont{Favata}},
  \bibinfo{author}{\bibfnamefont{K.~G.} \bibnamefont{Arun}}, \bibnamefont{and}
  \bibinfo{author}{\bibfnamefont{C.~K.} \bibnamefont{Mishra}},
  \bibinfo{journal}{Phys. Rev. D} \textbf{\bibinfo{volume}{93}},
  \bibinfo{pages}{124061} (\bibinfo{year}{2016}),
  \urlprefix\url{https://link.aps.org/doi/10.1103/PhysRevD.93.124061}.

\bibitem[{\citenamefont{Barack and Cutler}(2004)}]{PhysRevD.69.082005}
\bibinfo{author}{\bibfnamefont{L.}~\bibnamefont{Barack}} \bibnamefont{and}
  \bibinfo{author}{\bibfnamefont{C.}~\bibnamefont{Cutler}},
  \bibinfo{journal}{Phys. Rev. D} \textbf{\bibinfo{volume}{69}},
  \bibinfo{pages}{082005} (\bibinfo{year}{2004}),
  \urlprefix\url{https://link.aps.org/doi/10.1103/PhysRevD.69.082005}.

\bibitem[{\citenamefont{Cutler}(1998)}]{PhysRevD.57.7089}
\bibinfo{author}{\bibfnamefont{C.}~\bibnamefont{Cutler}},
  \bibinfo{journal}{Phys. Rev. D} \textbf{\bibinfo{volume}{57}},
  \bibinfo{pages}{7089} (\bibinfo{year}{1998}),
  \urlprefix\url{https://link.aps.org/doi/10.1103/PhysRevD.57.7089}.

\bibitem[{\citenamefont{Kocsis et~al.}(2007)\citenamefont{Kocsis, Haiman,
  Menou, and Frei}}]{PhysRevD.76.022003}
\bibinfo{author}{\bibfnamefont{B.}~\bibnamefont{Kocsis}},
  \bibinfo{author}{\bibfnamefont{Z.}~\bibnamefont{Haiman}},
  \bibinfo{author}{\bibfnamefont{K.}~\bibnamefont{Menou}}, \bibnamefont{and}
  \bibinfo{author}{\bibfnamefont{Z.}~\bibnamefont{Frei}},
  \bibinfo{journal}{Phys. Rev. D} \textbf{\bibinfo{volume}{76}},
  \bibinfo{pages}{022003} (\bibinfo{year}{2007}),
  \urlprefix\url{https://link.aps.org/doi/10.1103/PhysRevD.76.022003}.

\bibitem[{\citenamefont{Baibhav et~al.}(2020)\citenamefont{Baibhav, Berti, and
  Cardoso}}]{PhysRevD.101.084053}
\bibinfo{author}{\bibfnamefont{V.}~\bibnamefont{Baibhav}},
  \bibinfo{author}{\bibfnamefont{E.}~\bibnamefont{Berti}}, \bibnamefont{and}
  \bibinfo{author}{\bibfnamefont{V.}~\bibnamefont{Cardoso}},
  \bibinfo{journal}{Phys. Rev. D} \textbf{\bibinfo{volume}{101}},
  \bibinfo{pages}{084053} (\bibinfo{year}{2020}),
  \urlprefix\url{https://link.aps.org/doi/10.1103/PhysRevD.101.084053}.

\bibitem[{\citenamefont{Wang et~al.}(2020)\citenamefont{Wang, Ni, Han, Yang,
  and Zhong}}]{PhysRevD.102.024089}
\bibinfo{author}{\bibfnamefont{G.}~\bibnamefont{Wang}},
  \bibinfo{author}{\bibfnamefont{W.-T.} \bibnamefont{Ni}},
  \bibinfo{author}{\bibfnamefont{W.-B.} \bibnamefont{Han}},
  \bibinfo{author}{\bibfnamefont{S.-C.} \bibnamefont{Yang}}, \bibnamefont{and}
  \bibinfo{author}{\bibfnamefont{X.-Y.} \bibnamefont{Zhong}},
  \bibinfo{journal}{Phys. Rev. D} \textbf{\bibinfo{volume}{102}},
  \bibinfo{pages}{024089} (\bibinfo{year}{2020}),
  \urlprefix\url{https://link.aps.org/doi/10.1103/PhysRevD.102.024089}.

\bibitem[{\citenamefont{Ruan et~al.}(2021)\citenamefont{Ruan, Liu, Guo, Wu, and
  Cai}}]{Ruan2021}
\bibinfo{author}{\bibfnamefont{W.-H.} \bibnamefont{Ruan}},
  \bibinfo{author}{\bibfnamefont{C.}~\bibnamefont{Liu}},
  \bibinfo{author}{\bibfnamefont{Z.-K.} \bibnamefont{Guo}},
  \bibinfo{author}{\bibfnamefont{Y.-L.} \bibnamefont{Wu}}, \bibnamefont{and}
  \bibinfo{author}{\bibfnamefont{R.-G.} \bibnamefont{Cai}},
  \bibinfo{journal}{Research (Washington, D.C.)}
  \textbf{\bibinfo{volume}{2021}}, \bibinfo{pages}{6014164}
  (\bibinfo{year}{2021}), ISSN \bibinfo{issn}{2639-5274},
  \bibinfo{note}{33623919[pmid]},
  \urlprefix\url{https://pubmed.ncbi.nlm.nih.gov/33623919}.

\bibitem[{\citenamefont{Hu et~al.}(2021)\citenamefont{Hu, Li, Niu, and
  Zhao}}]{PhysRevD.103.064057}
\bibinfo{author}{\bibfnamefont{Q.}~\bibnamefont{Hu}},
  \bibinfo{author}{\bibfnamefont{M.}~\bibnamefont{Li}},
  \bibinfo{author}{\bibfnamefont{R.}~\bibnamefont{Niu}}, \bibnamefont{and}
  \bibinfo{author}{\bibfnamefont{W.}~\bibnamefont{Zhao}},
  \bibinfo{journal}{Phys. Rev. D} \textbf{\bibinfo{volume}{103}},
  \bibinfo{pages}{064057} (\bibinfo{year}{2021}),
  \urlprefix\url{https://link.aps.org/doi/10.1103/PhysRevD.103.064057}.

\bibitem[{\citenamefont{Zhang et~al.}(2021)\citenamefont{Zhang, Gong, Liu,
  Wang, and Zhang}}]{PhysRevD.103.103013}
\bibinfo{author}{\bibfnamefont{C.}~\bibnamefont{Zhang}},
  \bibinfo{author}{\bibfnamefont{Y.}~\bibnamefont{Gong}},
  \bibinfo{author}{\bibfnamefont{H.}~\bibnamefont{Liu}},
  \bibinfo{author}{\bibfnamefont{B.}~\bibnamefont{Wang}}, \bibnamefont{and}
  \bibinfo{author}{\bibfnamefont{C.}~\bibnamefont{Zhang}},
  \bibinfo{journal}{Phys. Rev. D} \textbf{\bibinfo{volume}{103}},
  \bibinfo{pages}{103013} (\bibinfo{year}{2021}),
  \urlprefix\url{https://link.aps.org/doi/10.1103/PhysRevD.103.103013}.

\bibitem[{\citenamefont{{Verbunt} and {Rappaport}}(1988)}]{1988ApJ...332..193V}
\bibinfo{author}{\bibfnamefont{F.}~\bibnamefont{{Verbunt}}} \bibnamefont{and}
  \bibinfo{author}{\bibfnamefont{S.}~\bibnamefont{{Rappaport}}},
  \bibinfo{journal}{The Astrophysical Journal} \textbf{\bibinfo{volume}{332}},
  \bibinfo{pages}{193} (\bibinfo{year}{1988}).

\bibitem[{\citenamefont{{Han} and {Webbink}}(1999)}]{1999A&A...349L..17H}
\bibinfo{author}{\bibfnamefont{Z.}~\bibnamefont{{Han}}} \bibnamefont{and}
  \bibinfo{author}{\bibfnamefont{R.~F.} \bibnamefont{{Webbink}}},
  \bibinfo{journal}{Astronomy \& Astrophysics} \textbf{\bibinfo{volume}{349}},
  \bibinfo{pages}{L17} (\bibinfo{year}{1999}).

\bibitem[{\citenamefont{{Yu} and {Jeffery}}(2010)}]{2010A&A...521A..85Y}
\bibinfo{author}{\bibfnamefont{S.}~\bibnamefont{{Yu}}} \bibnamefont{and}
  \bibinfo{author}{\bibfnamefont{C.~S.} \bibnamefont{{Jeffery}}},
  \bibinfo{journal}{Astronomy \& Astrophysics} \textbf{\bibinfo{volume}{521}},
  \bibinfo{eid}{A85} (\bibinfo{year}{2010}), \eprint{1007.4267}.

\bibitem[{\citenamefont{Seto}(2001)}]{PhysRevLett.87.251101}
\bibinfo{author}{\bibfnamefont{N.}~\bibnamefont{Seto}}, \bibinfo{journal}{Phys.
  Rev. Lett.} \textbf{\bibinfo{volume}{87}}, \bibinfo{pages}{251101}
  (\bibinfo{year}{2001}),
  \urlprefix\url{https://link.aps.org/doi/10.1103/PhysRevLett.87.251101}.

\bibitem[{\citenamefont{Willems et~al.}(2007)\citenamefont{Willems, Kalogera,
  Vecchio, Ivanova, Rasio, Fregeau, and Belczynski}}]{Willems_2007}
\bibinfo{author}{\bibfnamefont{B.}~\bibnamefont{Willems}},
  \bibinfo{author}{\bibfnamefont{V.}~\bibnamefont{Kalogera}},
  \bibinfo{author}{\bibfnamefont{A.}~\bibnamefont{Vecchio}},
  \bibinfo{author}{\bibfnamefont{N.}~\bibnamefont{Ivanova}},
  \bibinfo{author}{\bibfnamefont{F.~A.} \bibnamefont{Rasio}},
  \bibinfo{author}{\bibfnamefont{J.~M.} \bibnamefont{Fregeau}},
  \bibnamefont{and}
  \bibinfo{author}{\bibfnamefont{K.}~\bibnamefont{Belczynski}},
  \bibinfo{journal}{The Astrophysical Journal} \textbf{\bibinfo{volume}{665}},
  \bibinfo{pages}{L59} (\bibinfo{year}{2007}),
  \urlprefix\url{https://doi.org/10.1086/521049}.

\bibitem[{\citenamefont{Maggiore}(2008)}]{maggiore2008gravitational}
\bibinfo{author}{\bibfnamefont{M.}~\bibnamefont{Maggiore}},
  \emph{\bibinfo{title}{Gravitational waves: Volume 1: Theory and
  experiments}}, vol.~\bibinfo{volume}{1} (\bibinfo{publisher}{Oxford
  university press}, \bibinfo{year}{2008}).

\bibitem[{\citenamefont{Yunes et~al.}(2009)\citenamefont{Yunes, Arun, Berti,
  and Will}}]{PhysRevD.80.084001}
\bibinfo{author}{\bibfnamefont{N.}~\bibnamefont{Yunes}},
  \bibinfo{author}{\bibfnamefont{K.~G.} \bibnamefont{Arun}},
  \bibinfo{author}{\bibfnamefont{E.}~\bibnamefont{Berti}}, \bibnamefont{and}
  \bibinfo{author}{\bibfnamefont{C.~M.} \bibnamefont{Will}},
  \bibinfo{journal}{Phys. Rev. D} \textbf{\bibinfo{volume}{80}},
  \bibinfo{pages}{084001} (\bibinfo{year}{2009}),
  \urlprefix\url{https://link.aps.org/doi/10.1103/PhysRevD.80.084001}.

\bibitem[{\citenamefont{Tanay et~al.}(2016)\citenamefont{Tanay, Haney, and
  Gopakumar}}]{PhysRevD.93.064031}
\bibinfo{author}{\bibfnamefont{S.}~\bibnamefont{Tanay}},
  \bibinfo{author}{\bibfnamefont{M.}~\bibnamefont{Haney}}, \bibnamefont{and}
  \bibinfo{author}{\bibfnamefont{A.}~\bibnamefont{Gopakumar}},
  \bibinfo{journal}{Phys. Rev. D} \textbf{\bibinfo{volume}{93}},
  \bibinfo{pages}{064031} (\bibinfo{year}{2016}),
  \urlprefix\url{https://link.aps.org/doi/10.1103/PhysRevD.93.064031}.

\bibitem[{\citenamefont{Boetzel et~al.}(2017)\citenamefont{Boetzel, Susobhanan,
  Gopakumar, Klein, and Jetzer}}]{PhysRevD.96.044011}
\bibinfo{author}{\bibfnamefont{Y.}~\bibnamefont{Boetzel}},
  \bibinfo{author}{\bibfnamefont{A.}~\bibnamefont{Susobhanan}},
  \bibinfo{author}{\bibfnamefont{A.}~\bibnamefont{Gopakumar}},
  \bibinfo{author}{\bibfnamefont{A.}~\bibnamefont{Klein}}, \bibnamefont{and}
  \bibinfo{author}{\bibfnamefont{P.}~\bibnamefont{Jetzer}},
  \bibinfo{journal}{Phys. Rev. D} \textbf{\bibinfo{volume}{96}},
  \bibinfo{pages}{044011} (\bibinfo{year}{2017}),
  \urlprefix\url{https://link.aps.org/doi/10.1103/PhysRevD.96.044011}.

\bibitem[{\citenamefont{Stroeer et~al.}(2005)\citenamefont{Stroeer, Vecchio,
  and Nelemans}}]{Stroeer_2005}
\bibinfo{author}{\bibfnamefont{A.}~\bibnamefont{Stroeer}},
  \bibinfo{author}{\bibfnamefont{A.}~\bibnamefont{Vecchio}}, \bibnamefont{and}
  \bibinfo{author}{\bibfnamefont{G.}~\bibnamefont{Nelemans}},
  \bibinfo{journal}{The Astrophysical Journal} \textbf{\bibinfo{volume}{633}},
  \bibinfo{pages}{L33} (\bibinfo{year}{2005}),
  \urlprefix\url{https://doi.org/10.1086/498147}.

\bibitem[{\citenamefont{{Stroeer} and {Vecchio}}(2006)}]{2006CQGra..23S.809S}
\bibinfo{author}{\bibfnamefont{A.}~\bibnamefont{{Stroeer}}} \bibnamefont{and}
  \bibinfo{author}{\bibfnamefont{A.}~\bibnamefont{{Vecchio}}},
  \bibinfo{journal}{Classical and Quantum Gravity}
  \textbf{\bibinfo{volume}{23}}, \bibinfo{pages}{S809} (\bibinfo{year}{2006}),
  \eprint{astro-ph/0605227}.

\bibitem[{\citenamefont{Luo et~al.}(2016)\citenamefont{Luo, Chen, Duan, Gong,
  Hu, Ji, Liu, Mei, Milyukov, Sazhin et~al.}}]{luo2016tianqin}
\bibinfo{author}{\bibfnamefont{J.}~\bibnamefont{Luo}},
  \bibinfo{author}{\bibfnamefont{L.-S.} \bibnamefont{Chen}},
  \bibinfo{author}{\bibfnamefont{H.-Z.} \bibnamefont{Duan}},
  \bibinfo{author}{\bibfnamefont{Y.-G.} \bibnamefont{Gong}},
  \bibinfo{author}{\bibfnamefont{S.}~\bibnamefont{Hu}},
  \bibinfo{author}{\bibfnamefont{J.}~\bibnamefont{Ji}},
  \bibinfo{author}{\bibfnamefont{Q.}~\bibnamefont{Liu}},
  \bibinfo{author}{\bibfnamefont{J.}~\bibnamefont{Mei}},
  \bibinfo{author}{\bibfnamefont{V.}~\bibnamefont{Milyukov}},
  \bibinfo{author}{\bibfnamefont{M.}~\bibnamefont{Sazhin}},
  \bibnamefont{et~al.}, \bibinfo{journal}{Classical and Quantum Gravity}
  \textbf{\bibinfo{volume}{33}}, \bibinfo{pages}{035010}
  (\bibinfo{year}{2016}).

\bibitem[{\citenamefont{Apostolatos et~al.}(1994)\citenamefont{Apostolatos,
  Cutler, Sussman, and Thorne}}]{PhysRevD.49.6274}
\bibinfo{author}{\bibfnamefont{T.~A.} \bibnamefont{Apostolatos}},
  \bibinfo{author}{\bibfnamefont{C.}~\bibnamefont{Cutler}},
  \bibinfo{author}{\bibfnamefont{G.~J.} \bibnamefont{Sussman}},
  \bibnamefont{and} \bibinfo{author}{\bibfnamefont{K.~S.}
  \bibnamefont{Thorne}}, \bibinfo{journal}{Phys. Rev. D}
  \textbf{\bibinfo{volume}{49}}, \bibinfo{pages}{6274} (\bibinfo{year}{1994}),
  \urlprefix\url{https://link.aps.org/doi/10.1103/PhysRevD.49.6274}.

\bibitem[{\citenamefont{{Manchester}}(2015)}]{2015IJMPD..2430018M}
\bibinfo{author}{\bibfnamefont{R.~N.} \bibnamefont{{Manchester}}},
  \bibinfo{journal}{International Journal of Modern Physics D}
  \textbf{\bibinfo{volume}{24}}, \bibinfo{eid}{1530018} (\bibinfo{year}{2015}),
  \eprint{1502.05474}.

\bibitem[{\citenamefont{Ma and Yunes}(2019)}]{PhysRevD.100.124032}
\bibinfo{author}{\bibfnamefont{S.}~\bibnamefont{Ma}} \bibnamefont{and}
  \bibinfo{author}{\bibfnamefont{N.}~\bibnamefont{Yunes}},
  \bibinfo{journal}{Phys. Rev. D} \textbf{\bibinfo{volume}{100}},
  \bibinfo{pages}{124032} (\bibinfo{year}{2019}),
  \urlprefix\url{https://link.aps.org/doi/10.1103/PhysRevD.100.124032}.

\bibitem[{\citenamefont{Barausse et~al.}(2013)\citenamefont{Barausse,
  Palenzuela, Ponce, and Lehner}}]{PhysRevD.87.081506}
\bibinfo{author}{\bibfnamefont{E.}~\bibnamefont{Barausse}},
  \bibinfo{author}{\bibfnamefont{C.}~\bibnamefont{Palenzuela}},
  \bibinfo{author}{\bibfnamefont{M.}~\bibnamefont{Ponce}}, \bibnamefont{and}
  \bibinfo{author}{\bibfnamefont{L.}~\bibnamefont{Lehner}},
  \bibinfo{journal}{Phys. Rev. D} \textbf{\bibinfo{volume}{87}},
  \bibinfo{pages}{081506} (\bibinfo{year}{2013}),
  \urlprefix\url{https://link.aps.org/doi/10.1103/PhysRevD.87.081506}.

\bibitem[{\citenamefont{{Will} and {Zaglauer}}(1989)}]{1989ApJ...346..366W}
\bibinfo{author}{\bibfnamefont{C.~M.} \bibnamefont{{Will}}} \bibnamefont{and}
  \bibinfo{author}{\bibfnamefont{H.~W.} \bibnamefont{{Zaglauer}}},
  \bibinfo{journal}{The Astrophysical Journal} \textbf{\bibinfo{volume}{346}},
  \bibinfo{pages}{366} (\bibinfo{year}{1989}).

\bibitem[{\citenamefont{{Soffel} et~al.}(1987)\citenamefont{{Soffel}, {Ruder},
  and {Schneider}}}]{1987CeMec..40...77S}
\bibinfo{author}{\bibfnamefont{M.}~\bibnamefont{{Soffel}}},
  \bibinfo{author}{\bibfnamefont{H.}~\bibnamefont{{Ruder}}}, \bibnamefont{and}
  \bibinfo{author}{\bibfnamefont{M.}~\bibnamefont{{Schneider}}},
  \bibinfo{journal}{Celestial Mechanics} \textbf{\bibinfo{volume}{40}},
  \bibinfo{pages}{77} (\bibinfo{year}{1987}).

\bibitem[{\citenamefont{{Zhao} and {Xie}}(2013)}]{2013RAA....13.1231Z}
\bibinfo{author}{\bibfnamefont{S.-S.} \bibnamefont{{Zhao}}} \bibnamefont{and}
  \bibinfo{author}{\bibfnamefont{Y.}~\bibnamefont{{Xie}}},
  \bibinfo{journal}{Research in Astronomy and Astrophysics}
  \textbf{\bibinfo{volume}{13}}, \bibinfo{eid}{1231-1239}
  (\bibinfo{year}{2013}).

\bibitem[{\citenamefont{Damour et~al.}(1988)\citenamefont{Damour, Gibbons, and
  Taylor}}]{PhysRevLett.61.1151}
\bibinfo{author}{\bibfnamefont{T.}~\bibnamefont{Damour}},
  \bibinfo{author}{\bibfnamefont{G.~W.} \bibnamefont{Gibbons}},
  \bibnamefont{and} \bibinfo{author}{\bibfnamefont{J.~H.}
  \bibnamefont{Taylor}}, \bibinfo{journal}{Phys. Rev. Lett.}
  \textbf{\bibinfo{volume}{61}}, \bibinfo{pages}{1151} (\bibinfo{year}{1988}),
  \urlprefix\url{https://link.aps.org/doi/10.1103/PhysRevLett.61.1151}.

\bibitem[{\citenamefont{Barausse et~al.}(2016)\citenamefont{Barausse, Yunes,
  and Chamberlain}}]{PhysRevLett.116.241104}
\bibinfo{author}{\bibfnamefont{E.}~\bibnamefont{Barausse}},
  \bibinfo{author}{\bibfnamefont{N.}~\bibnamefont{Yunes}}, \bibnamefont{and}
  \bibinfo{author}{\bibfnamefont{K.}~\bibnamefont{Chamberlain}},
  \bibinfo{journal}{Phys. Rev. Lett.} \textbf{\bibinfo{volume}{116}},
  \bibinfo{pages}{241104} (\bibinfo{year}{2016}),
  \urlprefix\url{https://link.aps.org/doi/10.1103/PhysRevLett.116.241104}.

\bibitem[{\citenamefont{Zhao et~al.}(2021)\citenamefont{Zhao, Shao, Gao, Liu,
  Cao, and Ma}}]{PhysRevD.104.084008}
\bibinfo{author}{\bibfnamefont{J.}~\bibnamefont{Zhao}},
  \bibinfo{author}{\bibfnamefont{L.}~\bibnamefont{Shao}},
  \bibinfo{author}{\bibfnamefont{Y.}~\bibnamefont{Gao}},
  \bibinfo{author}{\bibfnamefont{C.}~\bibnamefont{Liu}},
  \bibinfo{author}{\bibfnamefont{Z.}~\bibnamefont{Cao}}, \bibnamefont{and}
  \bibinfo{author}{\bibfnamefont{B.-Q.} \bibnamefont{Ma}},
  \bibinfo{journal}{Phys. Rev. D} \textbf{\bibinfo{volume}{104}},
  \bibinfo{pages}{084008} (\bibinfo{year}{2021}),
  \urlprefix\url{https://link.aps.org/doi/10.1103/PhysRevD.104.084008}.

\bibitem[{\citenamefont{Damour and Esposito-Farese}(1992)}]{Damour_1992}
\bibinfo{author}{\bibfnamefont{T.}~\bibnamefont{Damour}} \bibnamefont{and}
  \bibinfo{author}{\bibfnamefont{G.}~\bibnamefont{Esposito-Farese}},
  \bibinfo{journal}{Classical and Quantum Gravity}
  \textbf{\bibinfo{volume}{9}}, \bibinfo{pages}{2093} (\bibinfo{year}{1992}),
  \urlprefix\url{https://doi.org/10.1088/0264-9381/9/9/015}.

\bibitem[{\citenamefont{Sch\"on and Doneva}(2022)}]{PhysRevD.105.064034}
\bibinfo{author}{\bibfnamefont{O.}~\bibnamefont{Sch\"on}} \bibnamefont{and}
  \bibinfo{author}{\bibfnamefont{D.~D.} \bibnamefont{Doneva}},
  \bibinfo{journal}{Phys. Rev. D} \textbf{\bibinfo{volume}{105}},
  \bibinfo{pages}{064034} (\bibinfo{year}{2022}),
  \urlprefix\url{https://link.aps.org/doi/10.1103/PhysRevD.105.064034}.

\bibitem[{\citenamefont{{Shakura}}(1985)}]{1985SvAL...11..224S}
\bibinfo{author}{\bibfnamefont{N.~I.} \bibnamefont{{Shakura}}},
  \bibinfo{journal}{Soviet Astronomy Letters} \textbf{\bibinfo{volume}{11}},
  \bibinfo{pages}{224} (\bibinfo{year}{1985}).

\bibitem[{\citenamefont{Kremer et~al.}(2015)\citenamefont{Kremer, Sepinsky, and
  Kalogera}}]{Kremer_2015}
\bibinfo{author}{\bibfnamefont{K.}~\bibnamefont{Kremer}},
  \bibinfo{author}{\bibfnamefont{J.}~\bibnamefont{Sepinsky}}, \bibnamefont{and}
  \bibinfo{author}{\bibfnamefont{V.}~\bibnamefont{Kalogera}},
  \bibinfo{journal}{The Astrophysical Journal} \textbf{\bibinfo{volume}{806}},
  \bibinfo{pages}{76} (\bibinfo{year}{2015}),
  \urlprefix\url{https://doi.org/10.1088/0004-637x/806/1/76}.

\bibitem[{\citenamefont{{Iben} et~al.}(1998)\citenamefont{{Iben}, {Tutukov},
  and {Fedorova}}}]{1998ApJ...503..344I}
\bibinfo{author}{\bibfnamefont{J.}~\bibnamefont{{Iben}}, \bibfnamefont{Icko}},
  \bibinfo{author}{\bibfnamefont{A.~V.} \bibnamefont{{Tutukov}}},
  \bibnamefont{and} \bibinfo{author}{\bibfnamefont{A.~V.}
  \bibnamefont{{Fedorova}}}, \bibinfo{journal}{Astrophys. J.}
  \textbf{\bibinfo{volume}{503}}, \bibinfo{pages}{344} (\bibinfo{year}{1998}).

\bibitem[{\citenamefont{{Willems} et~al.}(2010)\citenamefont{{Willems},
  {Deloye}, and {Kalogera}}}]{2010ApJ...713..239W}
\bibinfo{author}{\bibfnamefont{B.}~\bibnamefont{{Willems}}},
  \bibinfo{author}{\bibfnamefont{C.~J.} \bibnamefont{{Deloye}}},
  \bibnamefont{and}
  \bibinfo{author}{\bibfnamefont{V.}~\bibnamefont{{Kalogera}}},
  \bibinfo{journal}{Astrophys. J.} \textbf{\bibinfo{volume}{713}},
  \bibinfo{pages}{239} (\bibinfo{year}{2010}), \eprint{0904.1953}.

\bibitem[{\citenamefont{{Piro}}(2011)}]{2011ApJ...740L..53P}
\bibinfo{author}{\bibfnamefont{A.~L.} \bibnamefont{{Piro}}},
  \bibinfo{journal}{Astrophys. J. Lett.} \textbf{\bibinfo{volume}{740}},
  \bibinfo{eid}{L53} (\bibinfo{year}{2011}), \eprint{1108.3110}.

\bibitem[{\citenamefont{{Fuller} and {Lai}}(2012)}]{2012MNRAS.421..426F}
\bibinfo{author}{\bibfnamefont{J.}~\bibnamefont{{Fuller}}} \bibnamefont{and}
  \bibinfo{author}{\bibfnamefont{D.}~\bibnamefont{{Lai}}},
  \bibinfo{journal}{Mon. Not. R. Astron. Soc.} \textbf{\bibinfo{volume}{421}},
  \bibinfo{pages}{426} (\bibinfo{year}{2012}), \eprint{1108.4910}.

\bibitem[{\citenamefont{{Fuller} and {Lai}}(2013)}]{2013MNRAS.430..274F}
\bibinfo{author}{\bibfnamefont{J.}~\bibnamefont{{Fuller}}} \bibnamefont{and}
  \bibinfo{author}{\bibfnamefont{D.}~\bibnamefont{{Lai}}},
  \bibinfo{journal}{Mon. Not. R. Astron. Soc.} \textbf{\bibinfo{volume}{430}},
  \bibinfo{pages}{274} (\bibinfo{year}{2013}), \eprint{1211.0624}.

\bibitem[{\citenamefont{{Fuller} and {Lai}}(2014)}]{2014MNRAS.444.3488F}
\bibinfo{author}{\bibfnamefont{J.}~\bibnamefont{{Fuller}}} \bibnamefont{and}
  \bibinfo{author}{\bibfnamefont{D.}~\bibnamefont{{Lai}}},
  \bibinfo{journal}{Mon. Not. R. Astron. Soc.} \textbf{\bibinfo{volume}{444}},
  \bibinfo{pages}{3488} (\bibinfo{year}{2014}), \eprint{1406.2717}.

\bibitem[{\citenamefont{Thomas}(1977)}]{doi:10.1146.annurev.aa.15.090177.001015}
\bibinfo{author}{\bibfnamefont{H.-C.} \bibnamefont{Thomas}},
  \bibinfo{journal}{Annual Review of Astronomy and Astrophysics}
  \textbf{\bibinfo{volume}{15}}, \bibinfo{pages}{127} (\bibinfo{year}{1977}).

\bibitem[{\citenamefont{{Wong} and {Bildsten}}(2021)}]{2021ApJ...923..125W}
\bibinfo{author}{\bibfnamefont{T.~L.~S.} \bibnamefont{{Wong}}}
  \bibnamefont{and}
  \bibinfo{author}{\bibfnamefont{L.}~\bibnamefont{{Bildsten}}},
  \bibinfo{journal}{The Astrophysical Journal} \textbf{\bibinfo{volume}{923}},
  \bibinfo{eid}{125} (\bibinfo{year}{2021}), \eprint{2109.13403}.

\bibitem[{\citenamefont{Marsh et~al.}(2004)\citenamefont{Marsh, Nelemans, and
  Steeghs}}]{10.1111.j.1365-2966.2004.07564.x}
\bibinfo{author}{\bibfnamefont{T.~R.} \bibnamefont{Marsh}},
  \bibinfo{author}{\bibfnamefont{G.}~\bibnamefont{Nelemans}}, \bibnamefont{and}
  \bibinfo{author}{\bibfnamefont{D.}~\bibnamefont{Steeghs}},
  \bibinfo{journal}{Monthly Notices of the Royal Astronomical Society}
  \textbf{\bibinfo{volume}{350}}, \bibinfo{pages}{113} (\bibinfo{year}{2004}),
  ISSN \bibinfo{issn}{0035-8711},
  \eprint{https://academic.oup.com/mnras/article-pdf/350/1/113/3089404/350-1-113.pdf},
  \urlprefix\url{https://doi.org/10.1111/j.1365-2966.2004.07564.x}.

\bibitem[{\citenamefont{Gokhale et~al.}(2007)\citenamefont{Gokhale, Peng, and
  Frank}}]{Gokhale_2007}
\bibinfo{author}{\bibfnamefont{V.}~\bibnamefont{Gokhale}},
  \bibinfo{author}{\bibfnamefont{X.~M.} \bibnamefont{Peng}}, \bibnamefont{and}
  \bibinfo{author}{\bibfnamefont{J.}~\bibnamefont{Frank}},
  \bibinfo{journal}{The Astrophysical Journal} \textbf{\bibinfo{volume}{655}},
  \bibinfo{pages}{1010} (\bibinfo{year}{2007}),
  \urlprefix\url{https://doi.org/10.1086/510119}.

\bibitem[{\citenamefont{Guo}(2021)}]{10.3389.fspas.2021.663026}
\bibinfo{author}{\bibfnamefont{Z.}~\bibnamefont{Guo}},
  \bibinfo{journal}{Frontiers in Astronomy and Space Sciences}
  \textbf{\bibinfo{volume}{8}} (\bibinfo{year}{2021}), ISSN
  \bibinfo{issn}{2296-987X},
  \urlprefix\url{https://www.frontiersin.org/article/10.3389/fspas.2021.663026}.

\bibitem[{\citenamefont{Robson et~al.}(2019)\citenamefont{Robson, Cornish, and
  Liu}}]{Robson_2019}
\bibinfo{author}{\bibfnamefont{T.}~\bibnamefont{Robson}},
  \bibinfo{author}{\bibfnamefont{N.~J.} \bibnamefont{Cornish}},
  \bibnamefont{and} \bibinfo{author}{\bibfnamefont{C.}~\bibnamefont{Liu}},
  \bibinfo{journal}{Classical and Quantum Gravity}
  \textbf{\bibinfo{volume}{36}}, \bibinfo{pages}{105011}
  (\bibinfo{year}{2019}),
  \urlprefix\url{https://doi.org/10.1088%2F1361-6382%2Fab1101}.

\bibitem[{\citenamefont{Han et~al.}(2017)\citenamefont{Han, Cao, and
  Hu}}]{Han_2017}
\bibinfo{author}{\bibfnamefont{W.-B.} \bibnamefont{Han}},
  \bibinfo{author}{\bibfnamefont{Z.}~\bibnamefont{Cao}}, \bibnamefont{and}
  \bibinfo{author}{\bibfnamefont{Y.-M.} \bibnamefont{Hu}},
  \bibinfo{journal}{Classical and Quantum Gravity}
  \textbf{\bibinfo{volume}{34}}, \bibinfo{pages}{225010}
  (\bibinfo{year}{2017}),
  \urlprefix\url{https://doi.org/10.1088/1361-6382/aa891b}.

\bibitem[{\citenamefont{Sun et~al.}(2015)\citenamefont{Sun, Cao, Wang, and
  Yeh}}]{PhysRevD.92.044034}
\bibinfo{author}{\bibfnamefont{B.}~\bibnamefont{Sun}},
  \bibinfo{author}{\bibfnamefont{Z.}~\bibnamefont{Cao}},
  \bibinfo{author}{\bibfnamefont{Y.}~\bibnamefont{Wang}}, \bibnamefont{and}
  \bibinfo{author}{\bibfnamefont{H.-C.} \bibnamefont{Yeh}},
  \bibinfo{journal}{Phys. Rev. D} \textbf{\bibinfo{volume}{92}},
  \bibinfo{pages}{044034} (\bibinfo{year}{2015}),
  \urlprefix\url{https://link.aps.org/doi/10.1103/PhysRevD.92.044034}.

\bibitem[{\citenamefont{Ma et~al.}(2017)\citenamefont{Ma, Cao, Lin, Pan, and
  Yo}}]{PhysRevD.96.084046}
\bibinfo{author}{\bibfnamefont{S.}~\bibnamefont{Ma}},
  \bibinfo{author}{\bibfnamefont{Z.}~\bibnamefont{Cao}},
  \bibinfo{author}{\bibfnamefont{C.-Y.} \bibnamefont{Lin}},
  \bibinfo{author}{\bibfnamefont{H.-P.} \bibnamefont{Pan}}, \bibnamefont{and}
  \bibinfo{author}{\bibfnamefont{H.-J.} \bibnamefont{Yo}},
  \bibinfo{journal}{Phys. Rev. D} \textbf{\bibinfo{volume}{96}},
  \bibinfo{pages}{084046} (\bibinfo{year}{2017}),
  \urlprefix\url{https://link.aps.org/doi/10.1103/PhysRevD.96.084046}.

\bibitem[{\citenamefont{Pan et~al.}(2019)\citenamefont{Pan, Lin, Cao, and
  Yo}}]{PhysRevD.100.124003}
\bibinfo{author}{\bibfnamefont{H.-P.} \bibnamefont{Pan}},
  \bibinfo{author}{\bibfnamefont{C.-Y.} \bibnamefont{Lin}},
  \bibinfo{author}{\bibfnamefont{Z.}~\bibnamefont{Cao}}, \bibnamefont{and}
  \bibinfo{author}{\bibfnamefont{H.-J.} \bibnamefont{Yo}},
  \bibinfo{journal}{Phys. Rev. D} \textbf{\bibinfo{volume}{100}},
  \bibinfo{pages}{124003} (\bibinfo{year}{2019}),
  \urlprefix\url{https://link.aps.org/doi/10.1103/PhysRevD.100.124003}.

\bibitem[{\citenamefont{Yang et~al.}(2022)\citenamefont{Yang, Cai, Cao, and
  Lee}}]{PhysRevLett.129.191102}
\bibinfo{author}{\bibfnamefont{T.}~\bibnamefont{Yang}},
  \bibinfo{author}{\bibfnamefont{R.-G.} \bibnamefont{Cai}},
  \bibinfo{author}{\bibfnamefont{Z.}~\bibnamefont{Cao}}, \bibnamefont{and}
  \bibinfo{author}{\bibfnamefont{H.~M.} \bibnamefont{Lee}},
  \bibinfo{journal}{Phys. Rev. Lett.} \textbf{\bibinfo{volume}{129}},
  \bibinfo{pages}{191102} (\bibinfo{year}{2022}),
  \urlprefix\url{https://link.aps.org/doi/10.1103/PhysRevLett.129.191102}.

\bibitem[{\citenamefont{Yang et~al.}(2023)\citenamefont{Yang, Cai, Cao, and
  Lee}}]{PhysRevD.107.043539}
\bibinfo{author}{\bibfnamefont{T.}~\bibnamefont{Yang}},
  \bibinfo{author}{\bibfnamefont{R.-G.} \bibnamefont{Cai}},
  \bibinfo{author}{\bibfnamefont{Z.}~\bibnamefont{Cao}}, \bibnamefont{and}
  \bibinfo{author}{\bibfnamefont{H.~M.} \bibnamefont{Lee}},
  \bibinfo{journal}{Phys. Rev. D} \textbf{\bibinfo{volume}{107}},
  \bibinfo{pages}{043539} (\bibinfo{year}{2023}),
  \urlprefix\url{https://link.aps.org/doi/10.1103/PhysRevD.107.043539}.

\bibitem[{\citenamefont{{Romero-Shaw} et~al.}(2023)\citenamefont{{Romero-Shaw},
  {Gerosa}, and {Loutrel}}}]{2023MNRAS.519.5352R}
\bibinfo{author}{\bibfnamefont{I.~M.} \bibnamefont{{Romero-Shaw}}},
  \bibinfo{author}{\bibfnamefont{D.}~\bibnamefont{{Gerosa}}}, \bibnamefont{and}
  \bibinfo{author}{\bibfnamefont{N.}~\bibnamefont{{Loutrel}}},
  \bibinfo{journal}{Monthly Notices of the Royal Astronomical Society}
  \textbf{\bibinfo{volume}{519}}, \bibinfo{pages}{5352} (\bibinfo{year}{2023}),
  \eprint{2211.07528}.

\end{thebibliography}

\end{document}